\documentclass[a4paper,11pt]{article}
\usepackage{jcappub}

\usepackage{subfigure}
\usepackage{epsfig}
\usepackage{amsmath}
\usepackage{amsfonts}
\usepackage{amssymb}
\usepackage{cancel}
\usepackage{xfrac}
\usepackage{color}
\usepackage[utf8]{inputenc}
\usepackage{graphicx}
\usepackage{dcolumn}
\usepackage{bm}
\usepackage{tikz}
\usepackage{float}
\usepackage{tcolorbox}

\definecolor{rossos}{cmyk}{0,1,1,0.55}
\definecolor{bluscuro}{rgb}{0.15, 0.2, .85}
\definecolor{bluchiaro}{cmyk}{1,.3,0.,0.1}
\newcommand{\beq} {\begin{equation}}
\newcommand{\eeq} {\end{equation}}

\title{\boldmath Interacting fluid cosmologies from the metric-affine framework}

\author[a,1]{Fotios K. Anagnostopoulos,\note{Corresponding author.}}
\author[b,c]{Damianos Iosifidis}  

\affiliation[a]{Department of Informatics and Telecommunications, University of Peloponnese, Karaiskaki 70, 22100, Tripoli, Greece}
\affiliation[b]{Scuola Superiore Meridionale, Largo San Marcellino 10, 80138 Napoli, Italy}
\affiliation[c]{Istituto Nazionale di Fisica Nucleare – Sezione di Napoli, Via Cinthia, 80126 Napoli, Italy} 

\emailAdd{fotisanagn@uop.gr}
\emailAdd{fotis-anagnostopoulos@hotmail.com}
\emailAdd{d.iosifidis@ssmeridionale.it}	

\abstract{Metric-affine gravity provides a natural geometric framework in which spacetime curvature, torsion, and non-metricity are treated as independent degrees of freedom, leading to novel cosmological dynamics beyond General Relativity. A generic consequence of such theories is the emergence of effective interactions among the cosmological fluids, even in the absence of explicit phenomenological couplings. Interestingly, the above happens while the Standard Model of Particle physics remains unmodified. In this work, we investigate interacting cosmological models that arise from metric-affine gravity and analyze their implications for the background evolution of the Universe. We derive the modified continuity equations governing the matter, radiation, and dark-energy components, highlighting the geometric origin of energy exchange in the dark sector. We discuss how these interactions modify the expansion history and can effectively mimic evolving dark-energy behavior. The resulting cosmological scenarios are confronted with the most recent observational data from Type Ia Supernovae (Pantheon+), Cosmic Chronometers, and Baryon Acoustic Oscillations (DESI DR2). Moreover, using model selection criteria, e.g., Akaike Information Criterion, Bayesian Information Criterion and the Bayesian Evidence, we compare the aforementioned models against the standard $\Lambda$CDM model. We find that the proposed framework performs slightly better than $\Lambda CDM$, in the light of the data sets considered here. Our analysis demonstrates that metric-affine-induced interactions constitute a viable and theoretically motivated alternative for the description of the Universe.}

\begin{document}
\maketitle
\flushbottom

\section{Introduction} 
Cosmology today is at a crossroads, as the concordance model, i.e., the $\Lambda$ - Cold Dark Matter ($\Lambda$CDM) model, although able to adequately explain most cosmological observations, from Cosmic Chronometers (CC) \cite{moresco2018setting}, Type Ia Supernovae (SN Ia) \cite{rubin2025union}, Baryon Acoustic Oscillations (BAO) \cite{karim2025desi}, and the Cosmic Microwave Background (CMB) \cite{Aghanim:2018eyx}, suffers from the so-called cosmological tensions \cite{abdalla2022cosmology}. The most notable among them is the $H_0$ tension, namely the discrepancy between the value of the Hubble constant inferred under the assumption of $\Lambda$CDM from CMB observations and the ``model-independent'' value, fitted from late-universe astrophysical probes, such as Cepheid-calibrated distance ladders \cite{Riess_2019,Perivolaropoulos2024}. Also, there exist well-posed arguments from an epistemological point of view questioning the need for Dark Matter and Dark Energy entities \cite{merritt2017cosmology}, thus motivating a unified approach.

In parallel, the problem of a divergence-free quantum theory of gravity persists, despite the sustained efforts of the theoretical physics community. From an observational perspective, another central issue for the concordance model has recently emerged. Recent results from the Dark Energy Spectroscopic Instrument (DESI) have provided high-precision BAO measurements that, when analyzed in conjunction with CMB data and Type Ia supernova distance indicators, exhibit a mild but persistent inconsistency with the standard $\Lambda$CDM expansion history. These combined datasets favor departures from a strictly constant dark-energy component, suggesting a redshift-dependent behavior of the dark-energy sector. In the framework of commonly used phenomenological descriptions of an evolving equation of state, such as the Chevallier-Polarski-Linder (CPL) parametrization, the inferred deviation from $\Lambda$CDM reaches a statistical significance of approximately $2.8$-$3.8\sigma$ \cite{karim2025desi}.

In the context of addressing the $H_0$ tension, numerous modified cosmologies have appeared in the literature \cite{di2021realm}. A broad classification can be made based on their theoretical structure, distinguishing between cosmologies arising from modifications of General Relativity (GR) and phenomenological models. Among the well-studied cosmologies  in the former category are based on $f(R)$ \cite{de2010f}, $f(T)$ \cite{cai2016f,Capozziello:2011hj}, and $f(Q)$ modified gravities \cite{jimenez2018teleparallel}. However, it is known that the cases of $f(T)$ and $f(Q)$ are quite problematic, suffering a strong coupling problem \cite{BeltranJimenez:2021auj,Gomes:2023tur}. More viable models are constructed if, on top of curvature,  one adds either torsion \cite{kranas2019friedmann,Iosifidis:2021iuw} or nonmetricity \cite{Iosifidis:2022evi,Csillag:2024eor}, or both \cite{andrei2025friedmann}. Another particularly interesting class consists of scalar-tensor theories, with a notable example being Aether-Scalar-Tensor gravity \cite{PhysRevLett.127.161302}. The latter is capable of effectively describing astrophysical dark matter, thereby reducing the number of arbitrary entities required to model the Universe. On the other hand, the phenomenological class includes cosmologies that introduce \emph{ad hoc} modifications to the cosmological equations, such as the Friedmann and/or continuity equations. An exceptional member of this class is the so-called Running Vacuum cosmology \cite{basilakos2019scalar, gomez2023stringy, gomez2025composite}, which incorporates interactions in the dark sector and results in a modified cosmological evolution for both DM and radiation components. 

Interacting dark-sector models constitute a well-motivated extension of the standard cosmological paradigm, allowing for non-gravitational energy exchange between Dark Matter (DM) and Dark Energy (DE) \cite{Amendola2000, Bolotin2015}. Such interactions can naturally arise in effective field theory descriptions or scalar-tensor frameworks and have been extensively explored as a means to alleviate cosmological tensions, most notably the $H_0$ and $\sigma_8$ tensions \cite{di2021realm}. At the background level, interacting models modify the continuity equations, leading to altered expansion histories that can effectively mimic evolving dark-energy behavior without invoking explicit equation-of-state parametrizations \cite{Bolotin2015}. At the perturbative level, these interactions impact structure formation and growth-rate observables, providing additional discriminating power for observational tests \cite{Wang2016}. Several phenomenological coupling prescriptions, typically proportional to the DM or DE energy density, have been shown to yield competitive fits to CMB, BAO, and SN Ia datasets \cite{di2021realm}. Moreover, during the preparation of this work, a notable result appeared with regard to general interactive fluids \cite{Li:2026xaz} that are claimed to fit datasets significantly better than $\Lambda$CDM. 

As we mentioned above, there is a plethora of modified gravity theories that have a rather phenomenological character and lack a proper foundation. On the other hand, a geometrically well-motivated and fundamentally rigid exception is the Metric-Affine Gravity (MAG) framework \cite{hehl1995metric}. For various motives for studying MAG, see \cite{hehl1995metric} and references therein. In MAG, the Riemannian geometry constraint is released, and consequently, the connection is independent of the metric, allowing for the spacetime to possess torsion (the antisymmetric part of the connection) and non-metricity (the metric tensor is not covariantly constant). These new geometric features are associated with the micro-properties of matter \cite{hehl1995metric}. The microscopic characteristics of matter are described by the so-called hypermomentum tensor, which is formally defined by the variation of the matter sector of the theory with respect to the affine connection (see Sect. \ref{sect:Theory2} below). Another key characteristic of MAG is the fact that the usual energy-momentum conservation ceases to hold due to the presence of microstructure. These modified conservation laws offer a natural way to describe interacting fluids, where energy is pumped in and out of the system. It is exactly the aim of this work to show how these modifications can naturally describe interacting fluid cosmology and challenge the standard $\Lambda$CDM model.

This work is organized as follows: First, we present the basic geometric ingredients of MAG, also introducing its energy content. Then, in Sect. \ref{sect:Theory2}, we provide a concrete review of metric-affine cosmologies and present the corresponding modified Friedmann equations. In Sect. \ref{sect:observational}, we discuss the methods and datasets that we use to check the observational viability of the metric-affine scenarios, the corresponding results, and some implications. In Sect. \ref{sect:Discussion}, we discuss and comment on our findings while presenting other possible implications. Finally, in Sect. \ref{sect:Conclusion}, we outline our conclusions. 

\section{Geometric Setup and Matter Content}
\label{sect:Theory2}
We will work in a generalized Metric-Affine Geometry, that is, we consider a 4-dimensional manifold, which we endow with a metric $g$ and an independent affine connection $\nabla$. In local coordinates, their components read $g_{\mu\nu}$ and $\Gamma^{\lambda}{}_{\mu\nu}$ respectively. The connection is generic, and apart from curvature, it also admits torsion and non-metricity, which we define as
\begin{align}
R^{\mu}{}_{\nu\alpha\beta} &:=
2\partial_{[\alpha}\Gamma^{\mu} {}_{|\nu|\beta]}+2\Gamma^{\mu}{}_{\rho[\alpha}\Gamma^{\rho}{}_{|\nu|\beta]} \label{R},
\\
S_{\mu\nu}{}^{\lambda}&:=\Gamma^{\lambda}{}_{[\mu\nu]},
\\
Q_{\alpha\mu\nu} &:=- \nabla_{\alpha}g_{\mu\nu},
\end{align}
respectively. 
Now, the difference between the generic affine connection $\Gamma^{\lambda}{}_{\mu\nu}$ and the usual symmetric and metric-compatible Levi-Civita connection defines the so-called distortion tensor \cite{hehl1995metric}:
\begin{gather}
N^{\lambda}{}_{\mu\nu} :=\Gamma^{\lambda}{}_{\mu\nu}-\widetilde{\Gamma}^{\lambda}{}_{\mu\nu}= \nonumber \\
\frac{1}{2}g^{\alpha\lambda}(Q_{\mu\nu\alpha}+Q_{\nu\alpha\mu}-Q_{\alpha\mu\nu}) -g^{\alpha\lambda}(S_{\alpha\mu\nu}+S_{\alpha\nu\mu}-S_{\mu\nu\alpha}) \label{N}
\end{gather}
where $\widetilde{\Gamma}^{\lambda}{}_{\mu\nu}$ is the usual Levi-Civita connection derived solely from the metric and its first derivatives. Torsion and non-metricity can be retrieved from distortion according to \cite{hehl1995metric}:
\beq
S_{\mu\nu\alpha}=N_{\alpha[\mu\nu]} \, , 
\qquad 
Q_{\nu\alpha\mu}=2 N_{(\alpha\mu)\nu} \label{QNSN}
\eeq
These are geometric preliminaries. Now, we turn to the matter part. In MAG, in addition to the metrical energy-momentum tensor
\beq
T_{\mu\nu}:=-\frac{2}{\sqrt{-g}}\frac{\delta S_{M}}{\delta g^{\mu\nu}} \label{EMT}
\eeq
one also has the notion of hypermomentum, which is formally defined as the connection variational derivative of the matter action \cite{Hehl:1976hyperm,Hehl:1976kt,Hehl:1976kv}:
\beq
\Delta_{\lambda}{}^{\mu\nu}:=-\frac{2}{\sqrt{-g}}\frac{\delta S_{M}}{\delta \Gamma^{\lambda}{}_{\mu\nu}} \label{HMT}
\eeq
The energy-momentum tensor (\ref{EMT}) is the source of the metric field equations, whereas the hypermomentum tensor (\ref{HMT}) is the source of the connection field equations. Hypermomentum splits into its three physical pieces of spin, dilation, and shear according to \cite{Hehl:1976kt,hehl1995metric}:
\beq
\tau_{\mu\nu\alpha}:=\Delta_{[\mu\nu]\alpha} \;\;(spin) \label{s}
\eeq
\beq
\hat{\Delta}_{\mu\nu\alpha}=\Delta_{(\mu\nu)\alpha}-\frac{1}{n}\Delta_{\alpha} g_{\mu\nu} \;\;(shear)
\eeq
\beq
\Delta_{\alpha}:=\Delta_{\mu\nu\alpha}g^{\mu\nu} \;\;(dilation) \label{sh}
\eeq  
An extremely important consequence of the connection-matter couplings is the following: under diffeomorphisms, the connection part is also affected, and in particular, the diffeomorphism invariance of the matter sector yields the on-shell generalized conservation law\footnote{Or, better phrased, 'balance equation'. For the detailed derivation, see, for instance, \cite{Iosifidis:2020gth}.}
\begin{gather}
	\sqrt{-g}(2 \tilde{\nabla}_{\mu}T^{\mu}_{\;\;\alpha}-\Delta^{\lambda\mu\nu}R_{\lambda\mu\nu\alpha})+\hat{\nabla}_{\mu}\hat{\nabla}_{\nu}(\sqrt{-g}\Delta_{\alpha}^{\;\;\mu\nu})
    \nonumber \\
    +2S_{\mu\alpha}^{\;\;\;\;\lambda}\hat{\nabla}_{\nu}(\sqrt{-g}\Delta_{\lambda}^{\;\;\;\mu\nu})=0\label{ccc}
	\end{gather}
which is the generalization of the energy-momentum tensor conservation law for matter with microstructure. In the above, $ \hat{\nabla}_{\mu}:=2 S_{\mu}-\nabla_{\mu} $, and  $\tilde{\nabla}_{\mu}$ represents the covariant derivative computed with respect to the Levi-Civita connection. From (\ref{ccc}) we see that as long as the microstructure of matter is taken into account (i.e., $\Delta_{\lambda}{}^{\mu\nu}\neq 0$), the energy-momentum is not conserved. In the following, we will see how this fact naturally relates to the extended conservation laws that are frequently used in dynamical dark energy models. 

\subsection{Metric-Affine Cosmology}

We shall consider a flat FLRW Universe with the usual Robertson-Walker line element:
\beq
ds^{2}=-dt^{2}+a^{2}(t)\delta_{ij}dx^{i} dx^{j}
\eeq
where, as usual, $a(t)$ is the scale factor and $t$ is the cosmic time. In Metric-Affine Geometries, the independent connection with coefficients $\Gamma^{\lambda}{}_{\mu\nu}$ must also respect the underlying symmetries of homogeneity and isotropy. Spatial isotropy demands that the connection has a vanishing Lie derivative:
\beq
\mathcal{L}_{\xi}\Gamma^{\lambda}{}_{\mu\nu}=0
\eeq
whereas homogeneity allows only for time dependence for the connection coefficients. The above demand, when worked out \cite{Iosifidis:2020gth}, gives the form of the distortion for homogeneous and isotropic cosmologies as:
\begin{gather}
N_{\alpha\mu\nu}^{(n)}=X(t)u_{\alpha}h_{\mu\nu}+Y(t)u_{\mu}h_{\alpha\nu}+Z(t)u_{\nu}h_{\alpha\mu} \nonumber \\
	+V(t)u_{\alpha}u_{\mu}u_{\nu} +\epsilon_{\alpha\mu\nu\lambda}u^{\lambda}W(t)
	\end{gather}	
where the functions $X(t),Y(t),Z(t),V(t),W(t)$ represent the five degrees of freedom of the distortion. The five distortion degrees of freedom split into 2 degrees of freedom for torsion \cite{tsamparlis1979cosmological} and 3 for non-metricity \cite{minkevich1998isotropic}, and their corresponding covariant expressions read \cite{Iosifidis:2020gth}:
\beq
S_{\mu\nu\alpha}^{(n)}=2u_{[\mu}h_{\nu]\alpha}\Phi(t)+\epsilon_{\mu\nu\alpha\rho}u^{\rho}P(t)\delta_{n,4} \label{isotor}
\eeq
and 
	\beq
	Q_{\alpha \mu \nu}  = A(t) u_\alpha h_{\mu \nu} + B(t) h_{\alpha(\mu} u_{\nu)} + C(t) u_\alpha u_\mu u_\nu
	\eeq
The functions appearing in the last two equations relate to the distortion variables through:
\beq
2(X+Y)=B \;, \;\; 2Z=A\;, \;\; 2V=C \;, \;\; 2\Phi =Y-Z\;, \;\; P = W	 \label{dv}
\eeq
with inverse form:
\beq
W=P \;, \;\; V=C/2 \;, \;\; Z=A/2	
\eeq
\beq
Y=2\Phi +\frac{A}{2}	\;\;, \;\;\; X=\frac{B}{2}- 2 \Phi -\frac{A}{2}
\eeq
Up to now, we have discussed the geometrical part of the theory. Let us now focus on the matter part. There we have the two sources of MAG: the usual metrical energy-momentum tensor, $T_{\mu\nu}$, and the hypermomentum tensor \cite{Hehl:1976hyperm,Hehl:1976kt,Hehl:1976kv} $\Delta_{\lambda}{}^{\mu\nu}$. In fact, to be more precise, the canonical energy-momentum tensor along with hypermomentum are the true sources of MAG, and the metrical one is a byproduct of them. Of course, for FLRW Universes, the former has the usual perfect fluid form:
\beq
T_{\mu\nu}=\rho u_{\mu} u_{\nu}+p h_{\mu\nu} \label{T}
\eeq
where $h_{\mu\nu}:=g_{\mu\nu}+u_{\mu}u_{\nu}$ is the projection tensor. As for the latter, imposing isotropy and homogeneity demands that it must have the form (see \cite{Iosifidis:2020gth}):
\begin{gather}
\Delta_{\alpha\mu\nu}=2 \sigma u_{[\alpha}h_{\mu]\nu}+\epsilon_{\alpha\mu\nu\kappa}u^{\kappa}\zeta \nonumber \\+\Sigma_{2}\Big[ h_{\alpha\mu}+(n-1)u_{\alpha}u_{\mu} \Big] u_{\nu} +2 \Sigma_{1} u_{(\mu}h_{\alpha)\nu} \nonumber \\ +\frac{1}{4}\Delta u_{\alpha}g_{\mu\nu}
\end{gather}
where $\sigma(t),\zeta(t),\Sigma_{1}(t),\Sigma_{2}(t),\Delta(t)$ are the 5 hypermomentum functions allowed by the Cosmological Principle. The reader interested in an in-depth discussion of the Lagrangian formulation of Isotropic Hyperfluids and the associated thermodynamics is referred to \cite{Iosifidis:2023kyf}.
The first line above represents the spin contribution, the second the shears, and the last part is the dilation (see decomposition in equations (\ref{s})-(\ref{sh}) above). In this work, we will restrict ourselves to the pure spin case, and in particular the $\sigma$-part; namely, we shall take
\beq
\Delta_{\alpha\mu\nu}\equiv \tau_{\alpha\mu\nu}^{(\sigma)}=2 \sigma u_{[\alpha}h_{\mu]\nu} \label{purespin}
\eeq
as the source of the cosmological hyperfluid. \textcolor{black}{In this way, we isolate the role of spin, studying its cosmological effects without interference from the dilation and shear components of matter.}


The theory that we are going to analyze is given by:
\beq
S[g,\Gamma,\Phi]=\frac{1}{2 \kappa}\int d^{4}x\sqrt{-g}R+S_{M}
\eeq
where $R$ is the general scalar curvature constructed from the independent affine connection and $S_M$ is the matter sector of the theory, which we consider to be comprised of baryons, radiation, and a perfect pure spin hyperfluid, namely the collection (\ref{T}) and (\ref{purespin}) combined with the generalized conservation law (\ref{ccc}). For details on solving the connection field equation and the subsequent extraction of the Friedmann and continuity equations, see \cite{Andrei:2024vvy}. The associated system of equations for the FLRW cosmology then reads:
\begin{subequations}
\label{eq: spin only}
\begin{align}
\label{eq: FR1 spin only}
    3 H^{2} &= \kappa\rho + 3 H \kappa \sigma - \frac{3 \kappa^{2} \sigma^{2}}{4} \,, \\
    2 \dot{H} + 3 H^{2} &= - \kappa p + 2 H \kappa \sigma  + \dot{\sigma} \kappa - \frac{\kappa^{2} \sigma^{2}}{4} \,, \\
\label{eq: continuity spin only}
    \dot{\rho} + 3 H \left( \rho + p \right) &= \frac{\kappa \sigma}{2} \left( \rho + 3 p \right) \,,
\end{align}
\end{subequations}
\begin{align}
\label{eq: w_h spin only}
    w_{h}&=\frac{P_{h}}{\rho_{h}}= - \frac{8 H \sigma + 4 \dot{\sigma} - \kappa \sigma^{2}}{3 \sigma \left(4 H - \kappa \sigma\right)} \,.
\end{align}
where 
\begin{equation}
    \rho_{h} = 3H\sigma - 3\kappa\sigma^2/4
\end{equation}
and
\begin{equation}
    P_{h} = -2H\sigma - \dot{\sigma} +\kappa\sigma^2/4
\end{equation}
are, respectively, the effective density and pressure associated with the spin-hypermomentum contribution.

\subsection{Model formulation}
As already mentioned, as a first approach to model building, we restrict ourselves to the case with only spin components, i.e., $\Sigma_{i} =0$ and $\Delta=0$. Naturally, one might think that the first Friedmann equation contains no derivatives, making it a second-order algebraic equation that is easily solvable for the Hubble rate. However, this is not the case, as we have interacting fluids, thus we cannot decouple the components and solve the continuity equation independently per component. 

We assume that there exist three cosmic ingredients, i.e., $\rho_{} \equiv \rho_{tot} =\rho_{b+DM} + \rho_{rad} + \rho_{DE}$, where $\rho_{DE}$ describes Dark Energy that is in our case due to the hypermomentum of matter. That is, we identify the DE with the spin hypermomentum contribution:
\begin{equation}
    \rho_{DE} = \rho_{h}=3H\sigma - 3\kappa\sigma^2/4
\end{equation}
and
\begin{equation}
    P_{DE} =P_{h}= -2H\sigma - \dot{\sigma} +\kappa\sigma^2/4
\end{equation}
The above expressions have been extracted by comparing the two Friedmann equations with the standard form:
\begin{equation}
    3H^2 = \kappa \rho_{tot}
\end{equation}
and
\begin{equation}
    3H^2 + 2\dot{H}= -\kappa P_{tot}
\end{equation}
The first Friedmann equation reads as:
\begin{equation}
    E = \sqrt{\Omega_{DE} + \Omega_{r} + \Omega_{m}} \ ,
\end{equation}
where $\Omega_i :=\rho_i/\rho_{cr} = 8\pi G \rho_i/(3H_0^2)$ and $E := H/H_0$.

In standard cosmology, the equation-of-state (EoS) parameter \( w := p/\rho \) for baryonic matter and radiation is fixed by basic kinetic theory and relativistic thermodynamics. Baryonic matter at late times consists of non-relativistic particles, for which the pressure satisfies \( p = n k T \), while the energy density is dominated by the rest-mass contribution \( \rho \simeq n m c^2 \). As a consequence, \( p \ll \rho \), implying an effective equation of state \( w_{\mathrm{b}} \approx 0 \). Radiation, on the other hand, is composed of ultra-relativistic particles obeying the dispersion relation \( E = pc \). For an isotropic relativistic gas, the momentum flux yields a pressure \( p = \rho/3 \), leading to the fixed equation of state \( w_{\mathrm{r}} = 1/3 \). In standard General Relativity, the total energy-momentum tensor is covariantly conserved ($\nabla_\mu T^{\mu\nu}_{tot} = 0$). In the absence of direct energy exchange between cosmic fluids, this total conservation is linearly separated such that each fluid species is independently conserved. In our framework, the geometric modification induced by connection-matter couplings alter the conservation laws, leading to the generalized continuity equation \eqref{eq: continuity spin only} for the total fluid. Crucially, in order to maintain the integrity of the Standard Model of particle physics, we must assume that the fundamental fluid species (baryons, radiation, and the effective DE fluid) do not directly exchange energy with one another. Allowing for direct, macroscopic energy transfer between these sectors (e.g., standard matter decaying into the dark fluid) would require non-standard particle interactions that are heavily constrained by current local and cosmological observations. Therefore, we conceptualize the source term on the right-hand side of the continuity equation not as a phenomenological fluid-fluid interaction, but as a purely geometric effect, i.e., a universal coupling between the modified background spacetime and the matter sector. We interpret the source term in the modified conservation law not as a direct transfer of energy among the cosmic fluids, but as an exchange between each component and the metric-affine geometry. The decomposition \(T_{\rm tot}^{\mu\nu}=\sum_i T_i^{\mu\nu}\) ensures that the component balance equations must reproduce the total conservation law when summed; however, it does not uniquely determine how the geometric source term is distributed among the individual fluids. We therefore adopt, as a modeling assumption, a universal component-wise coupling to the geometry, while excluding explicit non-gravitational interactions among matter, radiation, and dark energy. Under this prescription, each component retains its standard equation of state but obeys a separate modified continuity equation governed by the same spin-hypermomentum variable:

\begin{equation}
    \dot{\rho}_{m}  + 3H\rho_{m} = \kappa \frac{\sigma}{2}\rho_{m}
\end{equation}

\begin{equation}
    \dot{\rho}_{r}  + 4H\rho_{r} = \kappa \sigma\rho_{r}
\end{equation}

\begin{equation}
    \dot{\rho}_{DE}  + 3H(1+w_{DE})\rho_{DE} = \kappa \frac{\sigma}{2}(1+3w_{DE})\rho_{DE}
\end{equation}
Note that for the case of $\sigma =0$ we retrieve standard non-interacting fluids, which corresponds to just GR without a cosmological constant. 

Dividing by $\rho_{cr}$ we have:
\begin{equation}
     \dot{\Omega}_{m}  + 3H\Omega_{m} = \kappa \frac{\sigma}{2}\Omega_{m}
\end{equation}

\begin{equation}
    \dot{\Omega}_{r}  + 4H\Omega_{r} = \kappa \sigma\Omega_{r}
\end{equation}

\begin{equation}
    \dot{\Omega}_{DE}  + 3H(1+w_{DE})\Omega_{DE} = \kappa \frac{\sigma}{2}(1+3 w_{D})\Omega_{DE}
\end{equation}

Going from time derivatives to redshift:
\begin{equation}
     -(1+z)\Omega'_{m}  + 3\Omega_{m} = \kappa \frac{\sigma}{2H}\Omega_{m} \label{Omegam}
\end{equation}

\begin{equation}
    -(1+z)\Omega'_{r}  + 4\Omega_{r} = \kappa \frac{\sigma}{H}\Omega_{r} \label{Omegar}
\end{equation}

\begin{equation}
     -(1+z)\Omega'_{DE}  + 3(1+w_{DE})\Omega_{DE} = \kappa \frac{\sigma}{2H}(1+3 w_{DE})\Omega_{DE}
\end{equation}
To proceed further, a functional form for the extra free function of the theory, $\sigma$, must be assumed.

\subsubsection{Cosmic density-related spin}
Here we assume that the squared spin is proportional to the total density of matter, viz.  $\sigma = b\sqrt{(\rho_b + \rho_r + \rho_{DE})/\kappa} = b \sqrt{3} H_0 E/\kappa$, where $b$ is a dimensionless parameter. This is supported by dimensional analysis and, in fact, in certain cases appears as an exact solution of the corresponding cosmological equations, see e.g. \cite{Iosifidis:2021iuw}. 
Under this assumption we have:
\begin{equation}
    \frac{\kappa \sigma}{2H} = \frac{\sqrt{3}b}{2E}\sqrt{\Omega_b + \Omega_r + \Omega_{DE}}=\frac{\sqrt{3}b}{2}
\end{equation}

\begin{equation}
    -(1+z)\frac{d\Omega_{DE}}{dz} = \Omega_{DE} \Big[ \frac{\sqrt{3}b}{2}-1 +\Big(\frac{3 \sqrt{3}b}{2}-1\Big)w_{DE} \Big]
\end{equation}

After some simple algebra, we obtain:
\begin{equation}
    w_{DE}=\frac{3b - 8 \sqrt{3}}{12 \sqrt{3}-9 b}+\frac{4}{3} \frac{E'}{E}\frac{1}{4-b\sqrt{3}}(1+z)
\end{equation}

$$w_{DE} = - \frac{8\sqrt{3}b - 3b^2 - 4\sqrt{3}b(1+z)E'E^{-1}}{3\sqrt{3}b(4 - \sqrt{3}b)}$$

and it follows that:
\begin{gather}
     -(1+z)\frac{d\Omega_{DE}}{dz} = \Omega_{DE} \Big[ \lambda_{1} +\lambda_{2}(1+z)\frac{\Omega_{DE}'+\Omega_{r}'+\Omega_{m}'}{(\Omega_{DE}+\Omega_{r}+\Omega_{m})}\Big] \label{Omegade}
\end{gather}
where:
\begin{equation}
    \lambda_{1}=-\sqrt{3}b+\frac{4 \sqrt{3}-6  b}{12 \sqrt{3}-9b}\;\;, \;\; \lambda_{2}=\frac{3 \sqrt{3}-2}{\kappa \sqrt{3}(4 \sqrt{3}-3b)}
\end{equation}

Now, (\ref{Omegam}) and (\ref{Omegar}) are directly integrable and give:
\begin{equation}
    \Omega_{m}=\Omega_{m,0}(1+z)^{3-\frac{\sqrt{3}b}{2}}
\end{equation}
and
\begin{equation}
    \Omega_{r}=\Omega_{r,0}(1+z)^{4-\sqrt{3}b}
\end{equation}
respectively. Substituting these into \eqref{Omegade}, we get the evolution equation:
\begin{gather}
     -\frac{d\Omega_{DE}}{dz} = \Omega_{DE}  \frac{\lambda_{1}}{(1+z)}  \nonumber \\
     +\Omega_{DE}\lambda_{2}\frac{\Omega_{DE}'+\Omega_{r,0}2(\tilde{b}-1)(1+z)^{2 \tilde{b}-3}+\Omega_{m,0}\tilde{b}(1+z)^{\tilde{b}-1}}{(\Omega_{DE}+\Omega_{r,0}(1+z)^{2(\tilde{b}-1)}+\Omega_{m,0}(1+z)^{\tilde{b}})}\label{Omegade}
\end{gather}
where $\tilde{b}=3-\sqrt{3}b/2$.
\begin{equation}
    \frac{d\Omega_{DE}}{dz} = - \frac{\Omega_{DE}}{1+z} \left[ \frac{ \lambda_1 + \lambda_2 E^{-2}\left(3 - \frac{\sqrt{3}b}{2}\right)\Omega_m + (4 - \sqrt{3}b)\Omega_r}{ 1 + \lambda_2 \Omega_{DE}E^{-2} } \right]
\end{equation}
where $$\lambda_1 = -\frac{4}{4 - \sqrt{3}b}, \ \lambda_2 = \frac{\sqrt{3}b - 2}{4 - \sqrt{3}b}.$$

The constant dependence of the deviance from the standard $\Lambda CDM$ is interesting. In particular, effective Dark Energy effects are turned on by a non-zero $b$, while a modified cosmic evolution for matter and radiation also turns on. So there are two competing factors: having enough DE to describe the late-time accelerated expansion while ensuring the modifications to the radiation evolution are small enough so as not to ruin the Big Bang Nucleosynthesis (BBN) constraints.  

In our model, the modified continuity equations dictate that the radiation energy density evolves as $\rho_r(z) = \rho_{r,0}(1+z)^{4-\sqrt{3}b}$. Assuming the photon fluid maintains a thermal blackbody spectrum, the Stefan-Boltzmann law requires $\rho_r \propto T^4$. Equating these expressions yields a modified temperature-redshift relation:
\begin{equation}
T(z) = T_0(1+z)^{1 - \frac{\sqrt{3}b}{4}}
\end{equation}
In the literature, deviations from the standard $\Lambda$CDM temperature evolution, $T_{\Lambda CDM}(z) = T_0(1+z)$, are typically parameterized as $T(z) = T_0(1+z)^{1-\beta}$. Our metric-affine framework therefore provides a direct geometric origin for this deviation, predicting $\beta = \frac{\sqrt{3}b}{4}$. However, the parameter $\beta$ is exceptionally well-constrained by multi-probe astrophysical observations. Measurements of the thermal Sunyaev-Zel'dovich (SZ) effect in galaxy clusters, combined with atomic and molecular absorption lines in high-redshift quasar spectra, strictly constrain the temperature evolution. For instance, recent comprehensive analyses yield $\beta = 0.012 \pm 0.016$ \cite{Luzzi2015} or similarly tight bounds $\vert{}\beta\vert{} \lesssim 0.03$ \cite{Avgoustidis2012, Hurier2014}. 

Taylor expanding for small $|b| \ll 1$ yields $w_{DE} \approx -2/3 + X/3 + \frac{\sqrt{3}}{12}b(X - 1)$, where the expansion factor $X \equiv (1+z)E'/E \ge 1$. Since $(X-1) \ge 0$, a positive $b \sim 10^{-3}$ pushes $w_{DE}$ further above $-1/3$. Conversely, a negative $b \sim -10^{-3}$ provides only a microscopic correction $\mathcal{O}(10^{-4})$, yielding $w_{DE} \approx -0.25$ today ($X \sim 1.23$). Consequently, the equation of state remains mathematically trapped above the $w_{DE} = -1/3$ threshold, strictly prohibiting late-time accelerated expansion unless early-universe bounds are violated.

\subsubsection{``Polytropic'' Spin model}

To overcome this fundamental limitation, it is necessary to decouple the late-time acceleration mechanism from the rigid early-universe constraints. This requires generalizing the interaction term to allow for more flexible evolutionary dynamics across cosmic history. To this end, we introduce a new free parameter, $n$, and define the interaction using a dimensionless coupling parameter $b$, such that it scales as $\frac{\kappa \sigma}{2H_0} = b E^n$, where $E = H/H_0 = \sqrt{\Omega_{tot}}$. In a less strict sense, this relation is analogous to the polytropic equation of state from \cite{andrei2025friedmann}, which served as an inspiration for the name. Such equations of state follow from statistical physics considerations. By injecting this generalized $\sigma \propto E^n$ assumption into the definition of the effective equation of state parameter $w_h$, and using the chain rule to convert $\dot{H}$ into a redshift derivative of the total energy density ($\Omega_{tot}'$), the equation of state separates cleanly:
\begin{equation}
\label{eq:w_h-model-ii}
w_{h} = - \frac{4 - b E^{n-1}}{3(2 - b E^{n-1})} + \frac{n(1+z)(\Omega_m' + \Omega_r' + \Omega_{DE}')}{3(2 - b E^{n-1})\Omega_{tot}}
\end{equation}

Substituting this $w_h$ back into the redshift continuity equation for Dark Energy and isolating $\Omega_{DE}'$ (where the prime denotes $d/dz$) causes all complex cross-terms to vanish. The system of ODEs for the evolution of energy densities reads as:
\begin{equation}
\frac{d\Omega_m}{dz} = \frac{\Omega_m}{1+z} \left( 3 - b E^{n-1} \right)
\end{equation}
\begin{equation}
\label{eq:m2-radiation-ode}
\frac{d\Omega_r}{dz} = \frac{\Omega_r}{1+z} \left( 4 - 2b E^{n-1} \right)
\end{equation}
\begin{equation} 
\frac{d\Omega_{DE}}{dz} = \frac{ \frac{\Omega_{DE} C_0}{1+z} + \Omega_{DE} C_1 \left( \frac{d\Omega_m}{dz} + \frac{d\Omega_r}{dz} \right) }{ 1 - \Omega_{DE} C_1 }
\end{equation}
where $$C_0 = \frac{2}{2 - b E^{n-1}}$$$$C_1 = \frac{n(1 - b E^{n-1})}{(2 - b E^{n-1})\Omega_{tot}}$$

To check the ability of the model to pass the Big Bang Nucleosynthesis constraints, we take an approximate form of eq. \eqref{eq:m2-radiation-ode}. At very high redshifts like BBN ($z \sim 10^9$), the universe is overwhelmingly dominated by radiation. This means the total energy density is almost exactly equal to the radiation density, i.e., $\Omega_{tot} \approx \Omega_r$.
Since $E = \sqrt{\Omega_{tot}}$, we can approximate $E \approx \sqrt{\Omega_r} = \Omega_r^{1/2}$. 
\begin{equation}
    \frac{d\Omega_r}{dz} \approx  \frac{\Omega_r}{1+z} \left( 4 - 2b\, \Omega_r^{\frac{n-1}{2}} \right)
\end{equation}
For $n > 1$, the exponent $(n-1)/2$ is strictly positive. Integrating backwards in cosmic time, as the radiation density grows, the modified interaction term $2b\, \Omega_r^{\frac{n-1}{2}}$ diverges. Consequently, at a critical high redshift, this term will inevitably reach a value of $4$, causing the bracket in the radiation evolution equation to vanish. Physically, this implies that the radiation density encounters an upper bound and abruptly ceases to scale with redshift. Furthermore, as $E^{n-1}$ grows, the denominator of the coupled dark energy equation ($2 - b E^{n-1}$) approaches zero, introducing a mathematical singularity that leads to a complete breakdown of the background ODE system. Thus, a model with $n > 1$ irreconcilably violates standard early-universe thermodynamics unless the coupling constant is fine-tuned to unphysically microscopic values (e.g., $b \sim 10^{-20}$), which would render the hypermomentum effects negligible at late times. Conversely, for an interaction index $n < 1$, the exponent $(n-1)/2$ is strictly negative. In the high-redshift limit, where $\Omega_r \gg 1$, the geometric interaction term scales inversely with the dominant energy density and is therefore heavily suppressed:
\begin{equation}
    2b\, \Omega_r^{\frac{n-1}{2}} \xrightarrow{z \to \infty} 0
\end{equation}

As a result, the geometric interaction effectively ``turns off'' dynamically in the early universe. The differential equation for radiation asymptotically reduces to its standard General Relativistic form:
$$\frac{d\Omega_r}{dz} \simeq - \frac{4\Omega_r}{1+z}$$
Therefore, imposing $n < 1$ acts as a natural regulatory mechanism. It restricts the spin-hypermomentum modification to be a purely late-time phenomenon, seamlessly recovering standard $\Lambda$CDM expansion during the BBN epoch without requiring extreme fine-tuning of the coupling parameter $b$. Notice though that the normalization condition implies that the parameters n,b are not independent.
Starting from
\begin{equation}
\rho_{\rm DE} = 3H\sigma - \frac{3\kappa\sigma^2}{4}, \qquad \frac{\kappa\sigma}{2H_0} = \beta E^n,
\end{equation}
the dimensionless geometric density is
\begin{equation}
\Omega_{\rm DE} = \frac{\kappa\rho_{\rm DE}}{3H_0^2} = 2\beta E^{n+1} - \beta^2E^{2n}.
\end{equation}
At the present epoch, $E(0)=1$, and hence
\begin{equation}
\Omega_{{\rm DE},0} = 2\beta - \beta^2.
\end{equation}
For a spatially flat universe,
\begin{equation}
1 = \Omega_{m0} + \Omega_{r0} + \Omega_{{\rm DE},0},
\end{equation}
so

\begin{equation}
\beta_\pm = 1 \pm \sqrt{\Omega_{m0} + \Omega_{r0}} 
\end{equation}
with
\begin{equation}
\Omega_{r0} = \frac{2.47 \times 10^{-5}}{h^2}.
\end{equation}
We employ the minus solution, which is the most general, as it avoids restricting the parameter space due to zero denominators in the quantities $C_0$ and $C_1$.

\subsection{Equivalent descriptions and relation to existing frameworks}
\label{subsec:equivalent_descriptions}

In this subsection, we clarify how the modified continuity equations, arising in our model, can be mapped onto several well-studied cosmological frameworks. Indeed, recall that in our geometric context, we have the generalized continuity equation
\begin{equation}
\dot{\rho}_i + 3H(1+w_i)\rho_i
= \frac{\kappa \sigma}{2}(1+3 w_{i})\rho_{i} ,
\label{eq:modified_continuity_equiv}
\end{equation}
for each cosmological fluid component \(i\), while preserving the standard equations of state \(w_m\simeq 0\) and \(w_r=1/3\). Given this fact, some comments are in order.

\paragraph{Particle creation cosmology.}
Equation \eqref{eq:modified_continuity_equiv} is mathematically identical to the continuity equation of cosmological particle production in an expanding universe \cite{lima2025cosmic},
\begin{equation}
\dot{\rho} + 3H(1+w_i)\rho = \Gamma \rho (1+w_i),
\end{equation}
with a particle creation/annihilation rate
\begin{equation}
\Gamma = \frac{\kappa \sigma}{2}\left( \frac{1+3 w_{i}}{1+w_{i}}\right) .
\end{equation}
With this identification, we have a clear interpretation of the spin hypermomentum function $\sigma$, namely, it is directly related to the particle production rate $\Gamma$. Such models were originally introduced within a non-equilibrium thermodynamic description of cosmology, where the Universe is treated as an open system exchanging energy with the gravitational field \cite{Prigogine1989,lima2025cosmic}. Note, however, that in the approach of \cite{lima2025cosmic}, each fluid density satisfies a continuity equation with a different $\Gamma$.

Applying our specific model formulations, if we assume the \emph{cosmic density-related spin} model where $\sigma = b\sqrt{3}H/\kappa$, the production rate reads:
\begin{equation}
\Gamma = \frac{b \sqrt{3}H}{2}\left( \frac{1+3 w_{i}}{1+w_{i}}\right) .
\end{equation}
Alternatively, under the generalized \emph{polytropic spin} model defined by the dimensionless coupling $\frac{\kappa \sigma}{2H_0} = b E^n$, the effective rate translates to:
\begin{equation}
\Gamma = b H_0 E^n \left( \frac{1+3 w_{i}}{1+w_{i}}\right) .
\end{equation}

\paragraph{Effective bulk viscous cosmology.}
An alternative but equivalent description can be formulated in terms of an effective bulk viscous pressure. Introducing a creation viscous pressure \(p_c\), the continuity equation can be rewritten as:
\begin{equation}
\dot{\rho} + 3H(\rho + p + p_c) = 0 ,
\end{equation}
with
\begin{equation}
p_c = -\frac{\Gamma}{3H}(\rho + p) .
\end{equation}
This corresponds to a bulk viscous fluid with an effective viscosity coefficient chosen such that the resulting background dynamics reproduces Eq.~\eqref{eq:modified_continuity_equiv}. While this mapping is exact at the homogeneous background level, differences generally arise at the level of linear cosmological perturbations.

\paragraph{Scalar-tensor theories and non-conserved matter.}
Modified conservation laws of the form \(\nabla_\mu T^{\mu\nu}\neq 0\) naturally appear in scalar-tensor theories formulated in the Jordan frame, where matter couples non-minimally to a scalar degree of freedom. In a homogeneous and isotropic background, one finds:
\begin{equation}
\dot{\rho} + 3H(\rho + p) = \beta(\phi)\dot{\phi}\,\rho .
\end{equation}
In the context of our \emph{cosmic density-related spin} model, this mathematical equivalence is achieved if the scalar field derivative evolves proportionally to the Hubble rate (\(\dot{\phi}\propto H\), meaning \(\phi\propto \ln(a)\)), allowing the equation to reduce to the form of Eq.~\eqref{eq:modified_continuity_equiv}. Conversely, for the \emph{polytropic spin} model, an equivalence requires a scalar field evolution obeying \(\dot{\phi} \propto H_0 E^n\). Hence, our geometric framework can be interpreted as an effective Jordan-frame description of a broader class of scalar-tensor theories with non-conserved matter \cite{Damour1993,Faraoni2004}.

\section{Observational aspects}
\label{sect:observational}
In this section, we discuss the observational consequences of the proposed polytropic spin model. Obviously, the complete study of the phenomenology goes well beyond the scope of this work. However, we perform detailed observational fits in order to \emph{compare} the aforementioned scenario with the concordance model in terms of its fitting quality. In addition, given the observational constraints on the free parameters, we discuss the implications of the polytropic spin model on the cosmic history.

\subsection{Data \& Methodology}
\label{sub:Data-Likelihood-analysis}
Here we present the methods we use in order (a) to constrain the free parameters of the model at hand, collectively expressed as $\phi^{\lambda}$, and (b) to assess the fitting quality of the model in comparison with the concordance model. We use late-universe data from Supernovae Ia (SNIa) observations, direct measurements of the Hubble function, i.e. Cosmic Chronometers (CC), and data from Baryon Acoustic Oscillations (BAO).

Given $S$ independent observational data-sets and assuming Gaussian errors, the total likelihood function is defined as:
\begin{equation}
        \mathcal{L}_{\text{tot}}(\phi^{\lambda}) \sim \prod_{p=1}^{\mathcal{P}} 
\exp[-\chi^2_{p}(\phi^{\psi})]\,,
\end{equation}
where the corresponding expression for $\chi_{\text{tot}}^2$ is written as:
\begin{equation}
        \chi_{\text{tot}}^2 = \sum_{p=1}^{\mathcal{P}}\chi^2_{p}\,.
\end{equation}
The statistical vector has a dimension of $k$, comprising $\nu$ parameters of the model under consideration plus $\nu_{i}$ \emph{intrinsic} parameters from the utilized data sets, yielding $k = \nu + \nu_{i}$. In the above expression, $\phi^{\mu}$ is the vector containing the free parameters. 

Lastly, to obtain the posterior distributions of the model parameters given the data, we utilize the Nested Sampling algorithm rather than traditional Markov Chain Monte Carlo (MCMC) techniques. This approach is highly advantageous as it simultaneously maps the posterior parameter space and computes the full multidimensional integral required for the Bayesian evidence. We perform the sampling using the open-source Python package \texttt{dynesty} \cite{speagle2020dynesty}, configuring the sampler with 1000 live points and utilizing the ``random'' slice sampling along all orientations (\texttt{rslice}) strategy to efficiently navigate complex, degenerate parameter contours. For assessing convergence, instead of relying on MCMC-specific diagnostics such as the Gelman-Rubin criterion or auto-correlation time, the nested sampling algorithm natively tracks the estimated remaining evidence. The run is dynamically terminated when this remaining evidence drops below a predefined tolerance threshold of $\Delta \ln \mathcal{Z} < 0.1$. The latter is a measure of the relative error at the final model selection assessment.

In the assessment of cosmological models based on their predictions with respect to the available data, we employ three widely recognized criteria: the Akaike Information Criterion (AIC), the Bayesian Information Criterion (BIC), and the Bayesian Evidence ($\mathcal{Z}$), i.e., \cite{liddle2007information} and references therein.

The AIC addresses the issue of model adequacy from an information theory perspective. Specifically, it serves as an estimator of the Kullback-Leibler information and possesses the property of asymptotic unbiasedness. Under the standard assumption of Gaussian posteriors, the modified AIC estimator is expressed as \cite{liddle2007information}:
\begin{equation}
\text{AIC}=-2\ln(\mathcal{L}_{\text{max}})+2k+\frac{2k(k+1)}{N_{\rm tot}-k-1}.
\end{equation}
Here, $\mathcal{L}_{\text{max}}$ represents the maximum likelihood of the considered data set(s), and $N_{\rm tot}$ is the total number of data points. For large $N_{\rm tot}$, the expression simplifies to $\text{AIC}\simeq -2\ln(\mathcal{L}_{\text{max}})+2k$, which corresponds to the standard form of the AIC. Consequently, it is considered good practice to utilize the modified AIC in all cases \cite{liddle2007information}.

The BIC is an asymptotic estimator of the Bayesian evidence, and its expression is given by:
\begin{equation}
\text{BIC} = -2\ln(\mathcal{L}_{\text{max}}) + k \cdot \log(N_{\text{tot}}).
\end{equation}
On the other hand, instead of relying on approximate information criteria, model selection can be performed robustly using the full Bayesian evidence, $\mathcal{Z}$. The evidence (or marginal likelihood) represents the probability of the data given the model, integrated over the entire parameter space:
$$\mathcal{Z} = \int \mathcal{L}(\phi^\mu) \pi(\phi^\mu) d\phi^\mu$$
where $\mathcal{L}(\phi^\mu)$ is the likelihood and $\pi(\phi^\mu)$ is the prior distribution of the parameters $\phi^\mu$. Unlike criteria that approximate the penalty for extra parameters using effective degrees of freedom, the Bayesian evidence inherently penalizes unwarranted model complexity (often referred to as the Bayesian Occam's razor) through the integration over the prior volume. To evaluate this computationally demanding multidimensional integral, we employ the Nested Sampling algorithm \cite{skilling2004nested}, which transforms the integral over the parameter space into a one-dimensional integral over the cumulative prior mass. Specifically, the evidence is computed numerically using \texttt{dynesty} \cite{speagle2020dynesty}, a dynamic nested sampling package that simultaneously estimates the posterior distributions and the full Bayesian evidence.

In the task of ranking a set of competing models based on their fitting quality to observational data, we employ the aforementioned criteria, specifically focusing on the relative difference in Information Criterion (IC) values within the given model set. The difference $\Delta \text{IC}_{\text{model}} = \text{IC}_{\text{model}} - \text{IC}_{\text{min}}$ allows us to assess the fitting quality between the competitive models using the Jeffreys scale \cite{Kass:1995loi}. In particular, a value of $\Delta\text{IC}\leq 2$ indicates the model is statistically compatible with the most favored model by the data, $2<\Delta\text{IC}<6$ implies a moderate tension between the two models, while $\Delta\text{IC}\geq 10$ indicates a significant tension.

\subsubsection{Supernovae Ia (SNIa)}

One of the primary objects regarding cosmological applications of standard candles in cosmology is the Type Ia supernova (SN Ia). We incorporate the full Pantheon+ sample \cite{scolnic2022pantheon+}, which expands upon the original Pantheon dataset by almost duplicating the sample size, improving the treatment of redshifts and peculiar velocities \cite{scolnic2018complete}, and greatly increasing the redshift range. The peculiar velocity treatment is particularly crucial for SNIa in the nearby universe, and relevant criticisms have been developed regarding the Pantheon sample \cite{colin2019evidence}. A key feature of the Pantheon+ dataset is the inclusion of multiple observations of the same SNe Ia from different surveys, with redundancy properly accounted for through an appropriate covariance matrix. The complete Pantheon+--SH0ES data product contains 1702 entries over the approximate redshift range ($0.00122\lesssim z\lesssim2.27$) \cite{PantheonPlusSH0ES_DataRelease}. However, we do not use the SNe Ia hosted by galaxies with Cepheid distance measurements that were employed by SH0ES to calibrate the absolute supernova luminosity scale. After removing these calibrator observations, our cosmological sample contains ($N_{\rm SN}=1624$) entries. The corresponding rows and columns are also removed from the full covariance matrix. Consequently, no Cepheid-based SH0ES calibration or local distance-ladder constraint is included in our likelihood. Instead, the SN Ia absolute magnitude parameter ($\mathcal{M}$) is treated as a free nuisance parameter and is constrained jointly with the cosmological parameters. The resulting supernova likelihood therefore uses only the relative distance-redshift information contained in the uncalibrated Pantheon+ Hubble diagram.

The chi-square function for this dataset is expressed as:
\begin{equation}
          \chi^{2}_{SN Ia}\left(\phi^{\nu}\right)={\bf{\delta \mu}}_{\text{SNIa}}\,
          {\bf C}_{\text{SNIa},\text{cov}}^{-1}\,{\bf{\delta \mu}}_{\text{SNIa}}^{T}\,,
\end{equation}
The vector ${\bf{\delta \mu}}_{\text{SNIa}}$ is defined as 
$  {\delta \mu}_{\text{SNIa}} = \{\mu_{1}-\mu_{\text{th}}(z_{1},\phi^{\nu})\,,\, ...\,, \, \mu_{N}-\mu_{\text{th}}(z_{N},\phi^{\nu})\}$. 
The distance modulus is defined as $\mu_{i} = \mu_{B,i}-\mathcal{M}$, where $\mu_{B,i}$ is the observed apparent magnitude at maximum in the rest frame for redshift $z_{i}$. 

The extra free parameter, $\mathcal{M}$, is essential to the analysis, as the observable distance modulus, $\mu_{obs}$, depends on modeling assumptions and, crucially, the fiducial cosmology \cite{scolnic2018complete}. Additionally, the theoretical form of the distance modulus is given by:
\begin{equation}
\mu_{\text{th}} = 5\log\left[\frac{d_{L}(z)}{\text{Mpc}}\right] + 25,
\end{equation}
with the luminosity distance defined as:
\begin{equation}
    d_L(z) = c(1+z)\int_{0}^{z} \frac{1}{H(\omega,\phi^{\nu})}d\omega,
\end{equation}

Note that $\mathcal{M}$ and the normalized Hubble constant $h$ exhibit intrinsic degeneracy within the context of the Pantheon dataset.  

\subsubsection{Cosmic Chronometers (CC)}

We compile and use a new dataset of H(z) data points, based on \cite{anagnostopoulos2024observational}. The aforementioned dataset has been built upon two considerations. First, we select data points that have been extracted via the differential age of passively evolving galaxies method (e.g., \cite{moresco2018setting, moresco2020setting} and references therein). This is done to minimize the dependency of the data points on the assumed fiducial cosmology. Indeed, this method is claimed to be almost unaffected by cosmological assumptions, except for the hypothesis of a flat FLRW geometry. Second, we consider the fact that the authors of Ref. \cite{ahlstrom2023use} were not able to reproduce the CC data points given by \cite{simon2005constraints}. For the benefit of the reader, we present the dataset in Tab. \ref{table:dataRatra}, along with the relevant references per data point.

\begin{table}
\centering
\begin{tabular}{ccccc}
\hline\hline
No. & z & $H \ [km/Mpc s^{-1}]$ & $\sigma [km/Mpc s^{-1}]$ & Ref.\\
\hline \hline
1 & 0.070 & 69.0 & 19.6 & \cite{zhang2014four}\\
2 & 0.120 & 68.6 & 26.2 & \cite{zhang2014four}\\
3 & 0.1791 & 75.0 & 4.0 & \cite{moresco2012improved}\\
4 & 0.1993 & 75.0 & 5.0 & \cite{moresco2012improved}\\
5 & 0.200 & 72.9 & 29.6 & \cite{zhang2014four} \\
6 & 0.280 & 88.8 & 36.6 & \cite{zhang2014four}\\
7 & 0.3519 & 83.0 & 14.0 & \cite{moresco2012improved}\\
8 & 0.3802 & 83.0 & 13.5 & \cite{moresco2016}\\
9 & 0.4004 & 77.0 & 10.2 & \cite{moresco2016}\\
10 & 0.4247 & 87.1 & 11.2 & \cite{moresco2016}\\
11 & 0.4497 & 92.8 & 12.9 & \cite{moresco2016}\\
12 & 0.47 & 89.0 & 50.0 & \cite{ratsimbazafy2017age}\\
13 & 0.4783 & 80.9 & 9.0 & \cite{moresco2016}\\
14 & 0.480 & 97.0 & 62.0 & \cite{stern2010cosmic}\\
15 & 0.5929 & 104.0 & 13.0 & \cite{moresco2012improved}\\
16 & 0.6797 & 92.0 & 8.0 & \cite{moresco2012improved}\\
17 & 0.7812 & 105.0 & 12.0 & \cite{moresco2012improved}\\
18 & 0.80 & 113.1 & 44.2 & \cite{jiao2023new} \\
19 & 0.8754 & 125.0 & 17.0 & \cite{moresco2012improved}\\
20 & 0.880 & 90.0 & 40.0 & \cite{stern2010cosmic}\\
21 & 1.037 & 154.0 & 20.0 & \cite{moresco2012improved}\\
22 & 1.26 & 135.0 & 65.0 & \cite{tomasetti2023new}\\
23 & 1.363 & 160.0 & 33.6 & \cite{moresco2015raising}\\
24 & 1.965 & 186.5 & 50.4 & \cite{moresco2015raising}\\ 
\hline\hline
\end{tabular}
\caption{The Cosmic Chronometers dataset used here. For details please see the relevan subsection in the text.}
\label{table:dataRatra}
\end{table}

Our analysis incorporates a total of $N_{CC}=24$ measurements of the Hubble rate, covering the redshift range $0.07 \lesssim z \lesssim 2$.

The corresponding $\chi^2_{H}$ function is expressed as:
\begin{equation}
          \chi^{2}_{CC}\left(\phi^{\nu}\right)={\bf \delta H}\,
          {\bf C}_{CC,\text{cov}}^{-1}\,{\bf \delta H}^{T}\,,
\end{equation}
with ${\bf \delta H}=\{H_{1}-H_{0}E(z_{1},\phi^{\nu})\,,\,...\,,\, H_{N}-H_{0}E(z_{N},\phi^{\nu})\}$, where $H_{i}$ are the observed Hubble rates at redshift $z_{i}$ ($i=1,...,N$). The covariance matrix has been defined at \cite{CCcovariance} according to \cite{moresco2020setting} and reads as:
\begin{equation}
    \label{eq:CC-cov}
     {\bf C}_{CC,\text{cov}} \equiv  {\rm C}^{\rm stat} + {\rm C}^{\rm met}+ {\rm C}^{\rm young} + {\rm C}^{\rm model}
\end{equation}
Some words regarding the origin of the aforementioned terms are in order. First, ${\rm C}^{\rm stat}$ is related to the correlations between the extracted data points due to the employed fitting procedure. Second, ${\rm C}^{\rm met}$ corresponds to the uncertainty in the stellar metallicity estimation. Third, the matrix ${\rm C}^{\rm young}$ is diagonal and represents the contribution to the error budget arising from the assumption that the observed galaxy spectrum is solely dominated by ``old" stars. Finally, the term ${\rm C}^{\rm model}$ describes the contribution to the total covariance matrix coming from modeling assumptions, which can be further decomposed as:
\begin{equation}
{\rm C}^{\rm model}={\rm C}^{\rm SFH}+{\rm C}^{\rm IMF}+{\rm C}^{\rm stelar \ lib.}+{\rm C}^{\rm SPS}
\end{equation}
where ${\rm C}^{\rm SFH}$ relates to the assumptions regarding the star formation history, ${\rm C}^{\rm IMF}$ relates to the initial mass function, ${\rm C}^{\rm stelar \ lib.}$ relates to the stellar library used, and finally, ${\rm C}^{\rm SPS}$ relates to the stellar population synthesis model used.

\subsubsection{Baryon Acoustic Oscillations}

\label{sect:Obs-baos}Baryon Acoustic Oscillations (BAO) are the result of acoustic waves propagating in the primordial photon-baryon plasma, driven by gravitational instabilities in the early Universe. Prior to recombination, i.e., before redshift $z \sim 1100$, baryons and photons were tightly coupled via Thomson scattering, and overdensities in the matter distribution led to outward-propagating sound waves. When the Universe cooled enough for electrons and protons to form neutral hydrogen, photons decoupled and streamed freely (producing the cosmic microwave background), while the baryon velocity field stalled. This left a residual overdensity at a characteristic co-moving scale—the sound horizon at the drag epoch, $r_d$—which is imprinted as a preferred separation scale in the late-time matter correlation function and galaxy clustering. BAO data thus provide a standard ruler that can be used to jointly constrain the Hubble expansion rate $H(z)$ and angular diameter distance $D_A(z)$, making them a powerful probe for DE, see e.g., \cite{eisenstein1998baryonic}. For our analysis, we specifically used the recent DESI DR2 BAO measurements \cite{abdul2025desiI, abdul2025desiII}. These data provide a dense sampling of the cosmic expansion history for the redshift span $0.1 < z < 4.2$. In order to prevent double-counting and systematic inconsistencies between survey pipelines, we rely exclusively on the DESI DR2 data rather than combining it with previous BAO datasets. This strict exclusion of legacy surveys inherently updates and replaces older discrete BAO collections, ensuring uniform calibration protocols across the entire matter correlation spectrum. The relevant $\chi^2_{\rm BAO}$ function is given by:
\begin{equation}\chi_{\rm BAO}^2(\phi^\nu) = {\bf y}  {\bf C}^{-1}{_\text{BAO}}  {\bf y}^T
\end{equation}
The residual vector ${\bf y} = \{y_0, y_1, \dots, y_{N-1}\}$ measures the difference between the theoretical prediction $\lambda_i(z_i, \phi^\nu)$ and the observed DESI DR2 measurement $d_{i, \text{obs}}$:
\begin{equation}y_i = \lambda_i(z_i, \phi^\nu) - d_{i, \text{obs}}, \quad i = 0, 1, \dots, N-1.
\end{equation}

\begin{table}[htbp]
\centering
\renewcommand{\arraystretch}{1.2}
\begin{tabular}{ccc}
\hline \hline
$z$ & Observable & Measurement \\ 
\hline 
0.295 & $D_V/r_d$ & 7.942 \\
0.510 & $D_M/r_d$ & 13.588 \\
0.510 & $D_H/r_d$ & 21.863 \\
0.706 & $D_M/r_d$ & 17.351 \\
0.706 & $D_H/r_d$ & 19.455 \\
0.934 & $D_M/r_d$ & 21.576 \\
0.934 & $D_H/r_d$ & 17.641 \\
1.321 & $D_M/r_d$ & 27.601 \\
1.321 & $D_H/r_d$ & 14.176 \\
1.484 & $D_M/r_d$ & 30.512 \\
1.484 & $D_H/r_d$ & 12.817 \\
2.330 & $D_H/r_d$ & 8.632 \\
2.330 & $D_M/r_d$ & 38.989 \\
\hline \hline
\end{tabular}
\caption{Baryon Acoustic Oscillations (BAO) measurements (rounded) from the DESI DR2 compilation. The dataset maps the cosmic expansion history via the isotropic ($D_V/r_d$), transverse comoving ($D_M/r_d$), and line-of-sight Hubble ($D_H/r_d$) distance observables across multiple redshift bins \cite{abdul2025desiI, abdul2025desiII}.}
\label{tab:desi_bao_data}
\end{table}

The covariance matrix for the BAO dataset is a sparse block-diagonal matrix, which can be expressed as:
\begin{equation}
\mathbf{C}_{\text{BAO}} = \text{diag}\left( C_0, C_1, C_2, C_3, C_4, C_5, C_6 \right),
\end{equation}
where the first element corresponds to the isotropic measurement and is given by a scalar $C_0 = 0.0058$. The remaining elements are $2 \times 2$ covariance blocks for the correlated $(D_M/r_d, D_H/r_d)$ measurements at each respective redshift bin (adopted from \cite{adame2025desi, abdul2025desiII} and rounded):
\begin{align*}
C_1 &= \begin{bmatrix} 0.0283 & -0.0326 \\ -0.0326 & 0.1839 \end{bmatrix}, \quad
C_2 = \begin{bmatrix} 0.0324 & -0.0237 \\ -0.0237 & 0.1115 \end{bmatrix}, \\[1ex]
C_3 &= \begin{bmatrix} 0.0262 & -0.0113 \\ -0.0113 & 0.0404 \end{bmatrix}, \quad
C_4 = \begin{bmatrix} 0.1053 & -0.0290 \\ -0.0290 & 0.0504 \end{bmatrix}, \\[1ex]
C_5 &= \begin{bmatrix} 0.5830 & -0.1952 \\ -0.1952 & 0.2683 \end{bmatrix}, \quad
C_6 = \begin{bmatrix} 0.0102 & -0.0231 \\ -0.0231 & 0.2827 \end{bmatrix}.
\end{align*}

Depending on the redshift bin and observation type, DESI DR2 reports isotropic ($D_V/r_d$), transverse comoving ($D_M/r_d$), or line-of-sight ($D_H/r_d$) BAO observables. Matching the exact dataset index structure, the theoretical observable vector $\lambda_i(z_i, \phi^\nu)$ is constructed as:

\begin{equation}
\label{eq:bao_indexing}\lambda_i(z_i, \phi^\nu) =
\begin{cases}\dfrac{D_V(z_i, \phi^\nu)}{r_d}, & \text{for } i = 0 \\[2.5ex]
\dfrac{D_M(z_i, \phi^\nu)}{r_d}, & \text{for } i \in {1, 3, 5, 7, 9, 12} \\[2.5ex]
\dfrac{D_H(z_i, \phi^\nu)}{r_d}, & \text{for } i \in {2, 4, 6, 8, 10, 11}\end{cases}
\end{equation}
Here, index $i=0$ corresponds to the low-redshift spherically averaged (isotropic) BAO measurement, while the remaining indices map the anisotropic decomposition into transverse comoving distances ($D_M/r_d$) and Hubble distances ($D_H/r_d$) across distinct redshift bins. The theoretical distance quantities entering Eq. (\ref{eq:bao_indexing}) are defined as follows:
\begin{subequations}
\begin{equation}
D_H(z, \phi^\nu) = \frac{c}{H(z, \phi^\nu)} 
\end{equation}
\begin{equation}
D_M(z, \phi^\nu) = (1+z) D_A(z, \phi^\nu) = \frac{c}{H_0} \int_{0}^{z} \frac{dz'}{E(z', \phi^\nu)},
\end{equation}
\begin{equation}
D_V(z, \phi^\nu) = \left[ \frac{c  z  D_A(z, \phi^\nu)^2 (1+z)^2}{H_0 E(z, \phi^\nu)} \right]^{1/3} 
\end{equation}
\end{subequations}
where $D_A(z, \phi^\nu) = d_L(z, \phi^\nu) / (1+z)^{2}$ is the angular diameter distance, $c$ is the speed of light in $\text{km/s}$, $H_0 = 100h \text{ km s}^{-1}\text{Mpc}^{-1}$, and $E(z, \phi^\nu) \equiv H(z, \phi^\nu)/H_0$ represents the dimensionless Hubble expansion rate.

The length scale \(r_d\) is the standard ruler, representing a characteristic scale of over-densities in the matter distribution. In the concordance \(\Lambda\)CDM model, BAO originate from sound waves in the early Universe, and \(r_d\) is equal to the comoving sound horizon \(r_s\) at the baryon drag epoch. However, for non-\(\Lambda\)CDM cosmological models, the origin of the standard ruler \(r_d\) might differ \cite{Verde:2016ccp}. In general, one may treat \(r_d\) as a free parameter rather than calculating it within the \(\Lambda\)CDM framework or marginalizing over it, for example, as in \cite{anagnostopoulos2019constraining} and references therein.

\subsection{Data analysis results}
\begin{table}
\centering

\begin{tabular}{cc}
\hline
\hline
Parameter & Prior \\
\hline
\hline
$\Omega_{m0}$ &$\mathcal{U}(0.001, 1.0) $\\
$n$ &$\mathcal{U}(-2.0, 0.95)$ \\
$h$ &$\mathcal{U}(0.60, 0.90)$ \\
$\mathcal{M}$ &$\mathcal{U}(-19.9, -18.0)$\\
$r_d/Mpc$ &$\mathcal{U}(110, 170)$\\
\hline
\hline
\end{tabular}
\caption{Priors used in the likelihood analysis.}
\label{tab:priors}
\end{table}
Using the data and the methodology described in Sect. \ref{sub:Data-Likelihood-analysis}, we sample the likelihood function for the considered full-model presented previously. Specifically, we employ the following free parameters/datasets compilations:
\begin{itemize}
    \item $\phi^{\mu} = 
\{\Omega_{m0},n,h, r_d\}$ and $\mathcal{P} = \{CC, BAOs\}$.  
\item $\phi^{\mu} = 
\{\Omega_{m0},n,h,\mathcal{M}, r_d\}$ and $\mathcal{P} = \{ SNIa, BAOs\}$.
\item $\phi^{\mu} = 
\{\Omega_{m0},n,h,\mathcal{M}, r_d\}$ and $\mathcal{P} = \{CC,SNIa, BAOs\}$.  
\end{itemize}

\begin{figure}
\includegraphics[scale=0.65]{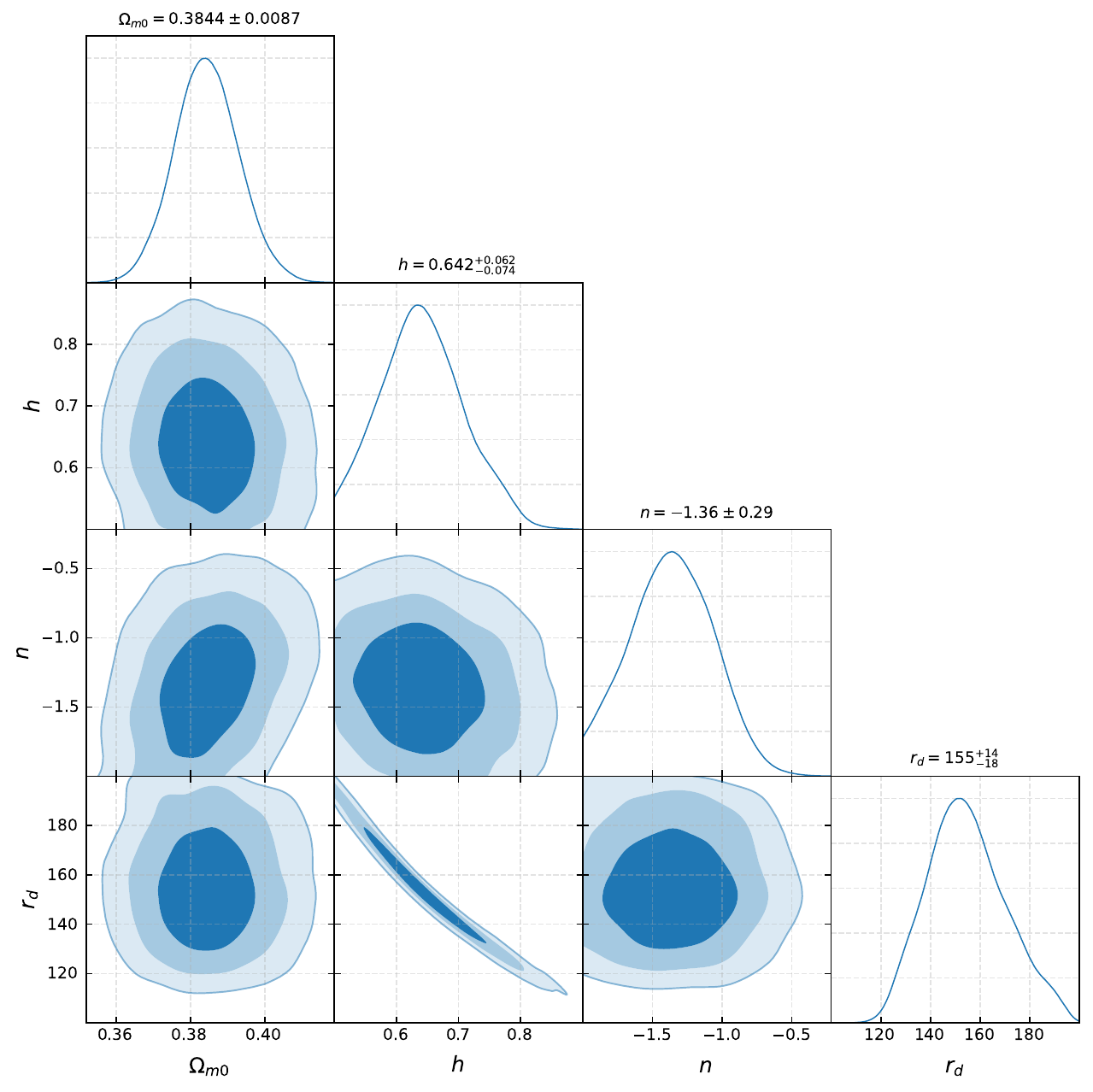} 
\caption[adf]{{\it{Iso-likelihood contours corresponding to quantile-defined $1\sigma-2\sigma-3\sigma$ areas for the case of CC+BAOs dataset.}}}
\label{fig:contours-noSNia} 
\end{figure}

\begin{figure}
\includegraphics[scale=0.65]{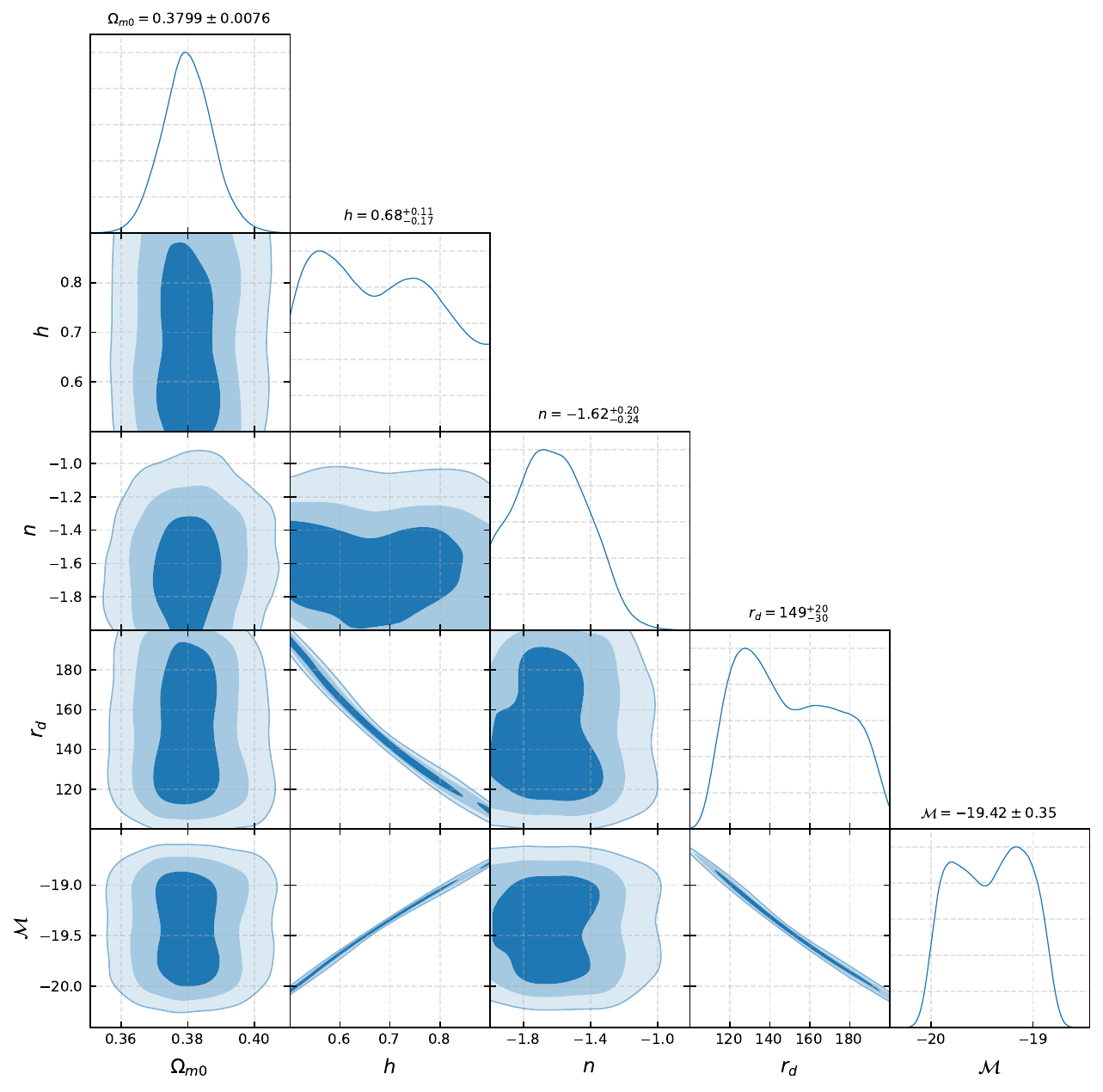} 
\caption[adf]{{\it{Iso-likelihood contours corresponding to quantile-defined $1\sigma-2\sigma-3\sigma$ areas for the case of SNIa/BAOs dataset.}}}
\label{fig:contours-SNIa-BAO} 
\end{figure}

\begin{figure}
\includegraphics[scale=0.65]{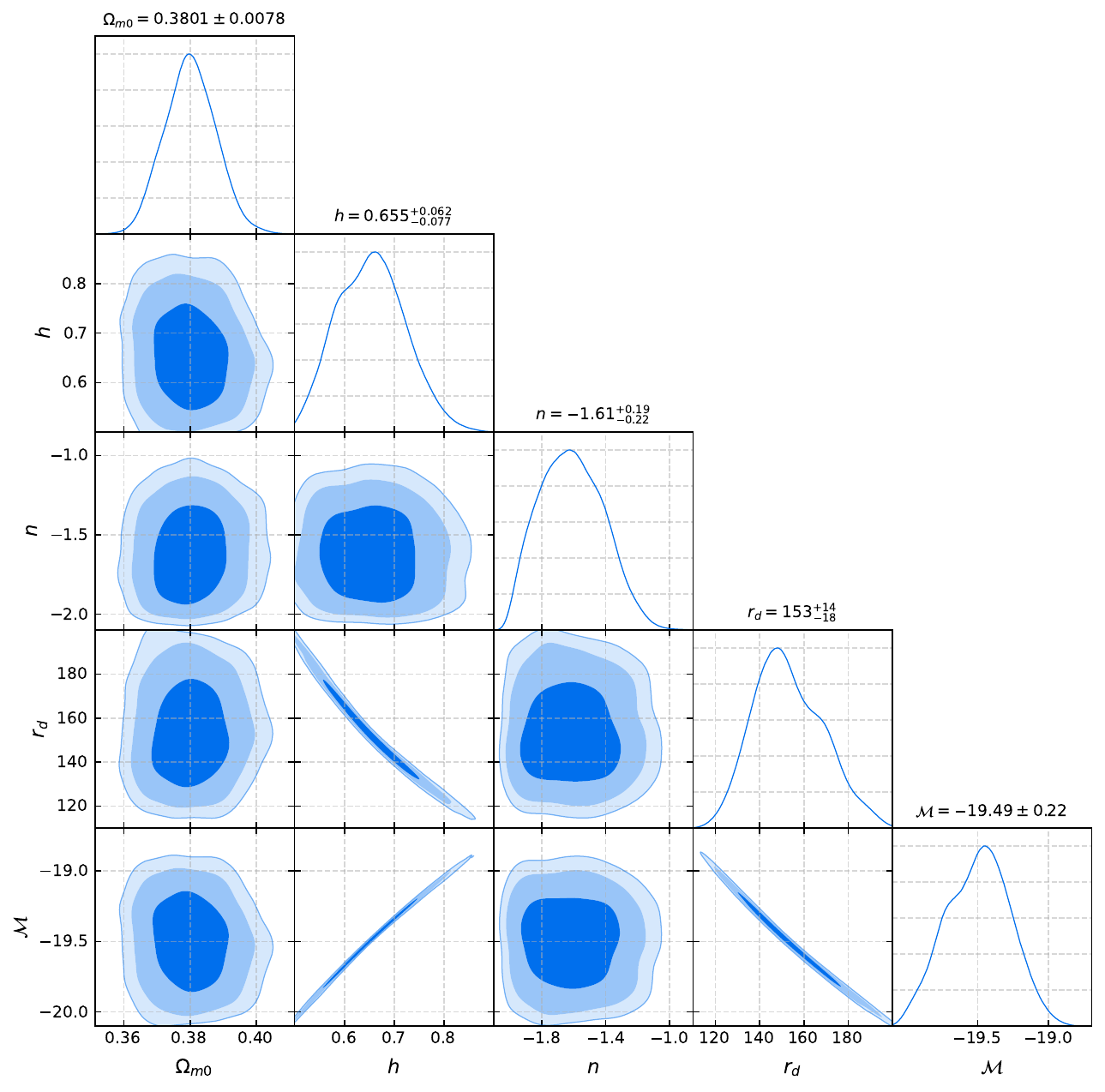} 
\caption[adf]{{\it{Iso-likelihood contours corresponding to quantile-defined $1\sigma-2\sigma-3\sigma$ areas for the case of SNIa/BAOs/CC dataset.}}}
\label{fig:contours-full} 
\end{figure}

\begin{figure}
\includegraphics[scale=0.65]{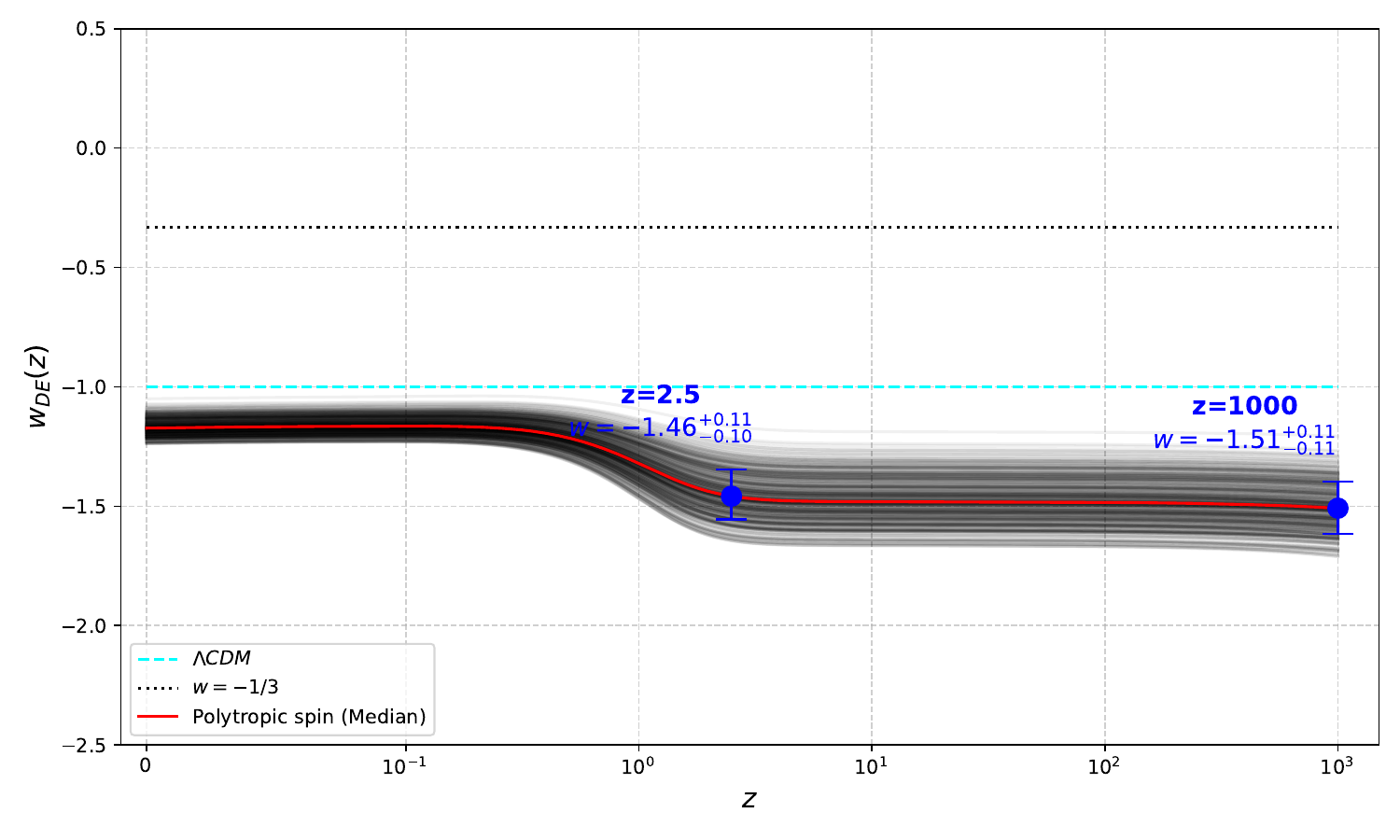} 
\caption[adf]{{\it{Plot of the equation of state parameter of the DE along the cosmic history. The x axis is in symmetric logarithmic scale. The black contour is made by a synthesis of $w_{DE}$ lines, constructed by resampling the MCMC chains for the case of CC+BAOs+SNIa dataset compilation. The red line corresponds to the best fit parameters, while the cyan is the $w = -1$ value.  The black dotted line corresponds to the $w_{DE} = -1/3$ case. The $w_{DE}(z=2.5)$ and $w_{DE}(z=1000)$ are also presented, along with the corresponding $1\sigma$ error bars.}}}
\label{fig:wde-reconstr} 
\end{figure}

We employ flat priors for the free parameters, as presented in Tab. \ref{tab:priors}. We depict 2D slices of the posterior distributions of the free parameter values for the polytropic spin model in Fig. \ref{fig:contours-noSNia}, \ref{fig:contours-SNIa-BAO}, and \ref{fig:contours-full} for the CC/BAOs, SNIa/BAOs, and CC/SNIa/BAOs dataset combinations, respectively. In Fig. \ref{fig:wde-reconstr}, we reconstruct the equation of state parameter for the polytropic spin model. 

In Tab. \ref{tab:parameters}, we summarize the parameter estimation results obtained for the proposed polytropic spin model compared against the standard flat $\Lambda$CDM cosmological baseline, along with their $1\sigma$ intervals. The latter have been calculated via quantiles on the posterior parameter distributions. Before assessing the physical implications of the results, some consistency checks are in order. We compare the $\Omega_{m0}$ values for the concordance $\Lambda$CDM model against the Planck 2018 results \cite{Aghanim:2018eyx} and report excellent $1\sigma$ compatibility across all dataset compilations considered, converging to $\Omega_{m0} \approx 0.3$ for all dataset compilations considered here. The same holds for the normalized Hubble constant, i.e., $h \sim 0.67$, and the BAO length scale, i.e., $r_d \sim 150$ Mpc.

Regarding the CC/BAOs dataset combination, we observe that the polytropic spin model favors a slightly larger matter density ($\Omega_{m0} = 0.384_{-0.008}^{+0.009}$) compared to the $\Lambda$CDM value. The extra free parameter, $n$, begin to show its influence here. The parameter  $n = -1.359_{-0.305}^{+0.303}$. Moving to the SNIa/BAOs combination, the constraints on the matter energy density for the polytropic spin model go to a little lower value ($\Omega_{m0} = 0.380 \pm 0.007$). Furthermore, the parameter $n$ decreases to $n = -1.6 \pm 0.2$, deviating from zero at approximately $8\sigma$ level. The absolute magnitude $\mathcal{M}$ stays fully consistent between both models in this configuration, yielding $\mathcal{M} = -19.44_{-0.41}^{+0.40}$ for the polytropic spin model and $\mathcal{M} = -19.40_{-0.47}^{+0.40}$ for $\Lambda$CDM. Note that exist a multimodality of the posterior distribution of h, which obviously appears also at $r_d$ and $\mathcal{M}$, as these quantities are  (positively/negatively) correlated.

Moreover, the joint analysis of all datasets (CC/SNIa/BAOs) confirms the intriguing trends observed in the partial dataset fits. The Hubble constant $h$ for the polytropic spin model is found to be slightly lower ($h = 0.659_{-0.065}^{+0.073}$) than its $\Lambda$CDM counterpart ($h = 0.674_{-0.075}^{+0.066}$), though both remain fully compatible with each other. Both the SNIa calibration parameter $\mathcal{M}$ and the present-day matter density remain stable. Note that the $\Omega_{m0}$ parameter for the polytropic spin model settles at almost $6.6\sigma$ above the $\Lambda CDM$ value.

For each criterion, we report the corresponding difference $\Delta \text{IC} \equiv \text{IC} - \text{IC}_{\text{min}}$ and $\Delta \ln \mathcal{Z} \equiv \ln \mathcal{Z}_{\text{max}} - \ln \mathcal{Z}$ to facilitate model comparison. For the CC/BAOs dataset combination, the Polytropic spin model is marginally favored by both the AIC ($\Delta \text{AIC} = 0.48$) and the Bayesian evidence ($\Delta \ln \mathcal{Z} = 0.20$), whereas $\Lambda$CDM is preferred under the BIC ($\Delta \text{BIC} = 1.13$). This trend becomes more pronounced when Supernovae Ia data are included (SNIa/BAOs and CC/SNIa/BAOs). In these combined datasets, the Polytropic model yields lower AIC values, achieving a superior fit with $\Delta \text{AIC} = 3.96$ and $\Delta \text{AIC} = 3.61$, respectively, over the concordance model. The Polytropic model is also favored by the log Bayesian evidence in both cases ($\Delta \ln \mathcal{Z} = 1.31$ and $1.32$). While $\Lambda$CDM remains favored under the BIC due to the penalization of the extra free parameters, the difference is noticeably narrower than in previous iterations ($\Delta \text{BIC} = 1.44$ and $1.77$). These metrics demonstrate that the metric-affine scenario performs highly competitively against $\Lambda$CDM across the employed datasets. We will discuss the physical implications of these results in detail in Sect. \ref{sect:Discussion}.
\begin{table*}[ht]
    \centering
    \caption{Parameter estimation results for the (-) branch of the Polytropic spin model. Results for the concordance model are also included in order to allow for direct comparison.}
    \label{tab:parameters}
    \begin{tabular}{cccccccc}
        \hline \hline
        Model & $\Omega_{m0}$ & $n$ & $h$ & $r_d$ & $\mathcal{M}$ & $\chi_{\text{min}}^{2}$ & $\chi_{\text{min}}/dof$ \\
        \hline
        \multicolumn{8}{c}{CC/BAOs} \\
        \hline
        Polytropic & $0.384_{-0.008}^{+0.009}$  & $-1.359_{-0.305}^{+0.303}$ & $0.639_{-0.065}^{+0.069}$ & $153.54_{-14.82}^{+17.73}$ & $-$ & $14.25$ & $0.43$ \\
       
        $\Lambda$CDM & $0.298_{-0.008}^{+0.008}$ & $-$ & $0.669_{-0.070}^{+0.074}$ & $151.60_{-15.30}^{+17.76}$ & $-$ & $16.73$ & $0.49$ \\
        \hline
        \multicolumn{8}{c}{SNIa/BAOs} \\
        \hline
        Polytropic & $0.380_{-0.007}^{+0.007}$ & $-1.631_{-0.212}^{+0.231}$ & $0.680_{-0.124}^{+0.130}$ & $146.04_{-23.73}^{+32.63}$ & $-19.40_{-0.44}^{+0.38}$ & $1464.69$ & $0.90$ \\
        
        $\Lambda$CDM & $0.305_{-0.008}^{+0.009}$ & $-$ &  $0.689_{-0.135}^{+0.137}$ & $146.51_{-24.43}^{+35.47}$ & $-19.40_{-0.47}^{+0.40}$ & $1470.65$ & $0.90$ \\
         \hline
        \multicolumn{8}{c}{{CC/SNIa/BAOs}} \\
        \hline
        
        Polytropic  & $0.380_{-0.008}^{+0.008}$ & $-1.620_{-0.206}^{+0.216}$ & $0.654_{-0.072}^{+0.070}$ & $151.42_{-14.17}^{+18.74}$ & $-19.49_{-0.25}^{+0.22}$ & $1471.42$ & $0.89$ \\
        
        $\Lambda$CDM & $0.305_{-0.008}^{+0.008}$ & $-$ & $0.674_{-0.075}^{+0.066}$ & $149.83_{-13.38}^{+18.39}$ & $-19.44_{-0.26}^{+0.20}$ & $1477.07$ & $0.89$ \\
        \hline
        \hline
    \end{tabular}
\end{table*}

\begin{table*}[ht]
\begin{center}
\tabcolsep 2.0pt
\vspace{1mm}
\begin{tabular}{ccccccc} \hline \hline
Model & AIC & $\Delta$AIC & BIC &$\Delta$BIC & $\ln \mathcal{Z}$ & $\Delta \ln \mathcal{Z}$
 \vspace{0.02cm}\\ 
 \hline
\hline

 \multicolumn{7}{c} {CC/BAOs}\\ 
  Polytropic spin model &  $22.25$ & $0.00$ & $28.69$ & $1.13$ & $-17.21$ & $0.00$ \\ 
   
$\Lambda$CDM & $22.73$ & $0.48$ & $27.56$ & $0.00$ & $-17.41$ & $0.20$ \\ 
  \vspace{0.05cm}\\ 
   \multicolumn{7}{c}{ {SNIa/BAOs}}\\
Polytropic spin model &  $1474.69$ & $0.00$ & $1501.69$ & $1.44$ & $-747.74$ & $0.00$  \\ 
   
$\Lambda$CDM & $1478.65$ & $3.96$  & $1500.25$ & $0.00$ & $-749.05$ & $1.31$\\
 \vspace{0.05cm}\\   
  \multicolumn{7}{c}{ {CC/SNIa/BAOs}}\\
Polytropic spin model &  $1481.46$ & $0.00$ & $1508.50$ & $1.77$ & $-751.92$ & $0.00$ \\
  
$\Lambda$CDM & $1485.07$ & $3.61$ & $1506.73$ & $0.00$ & $-753.24$ & $1.32$\\
 \hline\hline
  \vspace{0.02cm}\\ 
\end{tabular}
\caption{The information criteria 
AIC, BIC, and the Log Bayesian Evidence ($\ln \mathcal{Z}$) for the examined cosmological models,
alongside the corresponding differences 
$\Delta\text{IC} \equiv \text{IC} - \text{IC}_{\text{min}}$ and $\Delta \ln \mathcal{Z} \equiv \ln \mathcal{Z}_{\text{max}} - \ln \mathcal{Z}$.
\label{tab:Results2}}
\end{center}
\end{table*}

\section{Discussion}
\label{sect:Discussion}

The contrast between the two spin-hypermomentum parametrizations considered in this work clarifies the role played by the exponent \(n\). In the fixed \(n=1\) case, the dimensionless interaction strength \(\kappa\sigma/(2H)\) remains constant throughout the cosmic evolution. Consequently, the interaction cannot be important at late times without also modifying the matter and radiation sectors at high redshift, preventing an acceptable simultaneous description of the early- and late-time expansion histories. Allowing \(n\) to vary removes this rigidity. For the polytropic spin model,
\begin{equation}
\frac{\kappa\sigma}{2H_0}=bE^n,
\qquad
b=1-\sqrt{\Omega_{m0}+\Omega_{r0}},
\end{equation}
and hence
\begin{equation}
\frac{\kappa\sigma}{2H}=bE^{n-1}.
\end{equation}
The joint constraint \(n=-1.620^{+0.216}_{-0.206}\) implies a scaling approximately proportional to \(E^{-2.62}\). The interaction is therefore dynamically suppressed as the expansion rate increases, allowing matter and radiation to recover their standard background scalings toward the early universe while retaining a significant geometrical modification at late times.

This behavior represents the principal phenomenological advantage of the chosen \(\sigma(t)\) function. It separates the late-time acceleration mechanism from the high-redshift evolution without requiring an extremely small interaction amplitude. Nevertheless, the relation \(\sigma\propto E^n\) remains a phenomenological closure condition rather than a solution derived from a microscopic hyperfluid Lagrangian. Although the modified continuity equations follow from the metric-affine conservation laws, the particular time dependence assigned to the spin variable must ultimately be justified through a dynamical theory of the hyperfluid.

A characteristic result of the viable minus branch is its preference for a larger present-day matter abundance, 
$
\Omega_{m0}\simeq0.38,
$ compared with \(\Omega_{m0}\simeq0.305\) in \(\Lambda\)CDM. This should not be interpreted as a reduction in the amount of physical dark matter required by the model. Under spatial flatness, the inferred matter abundance instead corresponds to a smaller present-day effective dark-energy density, \(\Omega_{{\rm DE},0}\simeq0.62\). The larger matter contribution raises \(E(z)\) at the intermediate redshifts probed by BAO and compensates for the decreasing importance of the geometric component toward the past.

It would be premature to regard \(\Omega_{m0}\simeq0.38\) as incompatible with the CMB solely because it differs from the value inferred within \(\Lambda\)CDM. The acoustic peaks constrain combinations of background and perturbation quantities within a specified cosmological framework, and the inferred matter abundance is generally model dependent when the dark-sector conservation equations are modified. A relevant example is provided by Ref.~\cite{Li:2026xaz}, which analyzes coupled-quintessence and coupled-fluid interacting-dark-energy models using CMB temperature, polarization, and lensing measurements from Planck, ACT, and SPT, together with DESI DR2 BAO and several Type Ia supernova compilations. That study reports a \(3\!-\!5\sigma\) preference for a non-vanishing interaction and finds that the interacting models can match or improve upon the fits obtained with \(\Lambda\)CDM and the CPL parametrization.

In particular, the coupled-fluid realization of Ref.~\cite{Li:2026xaz} favors \(\Omega_{m0}\simeq0.59\!-\!0.63\), depending on the dataset combination. The large matter abundance is generated by an energy transfer from dark energy to dark matter and is compensated at the background level by a strongly phantom dark-energy equation of state. By contrast, the coupled-quintessence realization in the same analysis retains \(\Omega_{m0}\simeq0.30\). A large value of \(\Omega_{m0}\) is therefore not a generic prediction of interacting cosmology but a model-dependent consequence of the interaction law. The comparison demonstrates that matter abundances appreciably above the \(\Lambda\)CDM inference can remain compatible with full CMB spectra, but it does not establish the CMB viability of the present metric-affine model.

The reconstructed effective equation of state provides the complementary part of the physical interpretation. From Eq.~\eqref{eq:w_h-model-ii}, the joint-fit median parameters give approximately:

$$
w_{\rm DE}(0)\simeq-1.17,\qquad
w_{\rm DE}(2.5)\simeq-1.45,\qquad
w_{\rm DE}(1000)\simeq-1.51.
$$

\noindent The geometric component is therefore phantom-like. Its negative effective pressure compensates for the larger matter abundance and enables the model to reproduce the observed distance--redshift relation. This compensation is qualitatively reminiscent of the high-\(\Omega_{m0}\) coupled-fluid solution of Ref.~\cite{Li:2026xaz}, although the two models differ in both their theoretical origin and their energy-transfer mechanism.

Because the dark-energy contribution is constructed from the metric-affine geometry, \(w_{\rm DE}<-1\) does not by itself imply the existence of a fundamental phantom scalar field or a ghost degree of freedom. The equation of state characterizes an effective background fluid and cannot determine the stability of the underlying gravitational theory. Moreover, the suppression of \(bE^{n-1}\) at high redshift is accompanied by a decreasing fractional contribution of the geometric sector toward the past. The model is therefore more appropriately interpreted as a late-time geometric dark-energy scenario than as a conventional Early Dark Energy injection.

The statistical comparison supports this interpretation without providing decisive model selection. For CC/BAO, the AIC difference is only \(\Delta{\rm AIC}=0.48\) in favor of the polytropic model, while the BIC favors \(\Lambda\)CDM by \(\Delta{\rm BIC}=1.13\). The corresponding relative log Bayesian evidence is \(\Delta\ln\mathcal{Z}=0.20\) in favor of the polytropic model, rendering the two scenarios effectively indistinguishable for this dataset combination.

When SNe Ia are included, the improvement in the minimum \(\chi^2\) is sufficient for the AIC to favor the polytropic model. Relative to \(\Lambda\)CDM, the AIC improvements are \(\Delta{\rm AIC}=3.96\) for SNIa/BAO and \(\Delta{\rm AIC}=3.61\) for CC/SNIa/BAO. The BIC instead mildly favors \(\Lambda\)CDM by \(1.44\) and \(1.77\), respectively, because it applies a stronger penalty to the additional parameter. The Bayesian evidence favors the polytropic model by \(\Delta\ln\mathcal{Z}=1.31\) and \(1.32\), corresponding to Bayes factors of approximately \(3.7\). These values represent weak positive support rather than decisive evidence. The improvement in likelihood is therefore sufficient to make the polytropic spin model statistically competitive with \(\Lambda\)CDM, but not to establish a clear observational preference. The evidence should also be tested against reasonable extensions of the prior on \(n\), given the relative proximity of the posterior to its adopted lower boundary.

At the homogeneous-background level, the modified conservation equations admit algebraic mappings onto particle-creation, bulk-viscous, interacting-dark-energy, and particular Jordan-frame scalar-tensor cosmologies. A complementary comparison is provided by Ref.~\cite{deCruzPerez:2025dni}, which confronts several dynamical-dark-energy and running-vacuum scenarios with Pantheon+ or DES-Y5 supernovae, DESI DR2 BAO, and Planck PR4 CMB data. In those models, the cosmological evolution is specified through an evolving dark-energy equation of state, a vacuum density depending on the Hubble rate, or an explicit interaction between vacuum energy and dark matter. The metric-affine interpretation is conceptually different. The cosmic fluids are not assumed to exchange energy directly with one another. Instead, their modified evolution originates from an exchange with the non-Riemannian geometry. The geometrical sector therefore acts as the mediator of the effective interaction. This distinction may remain hidden in the homogeneous expansion history, where theories with different microscopic foundations can produce nearly degenerate predictions, but it can become observable at the perturbative level.

A complete assessment of the model consequently requires the derivation of the perturbed metric-affine field equations and the associated perturbed conservation laws. This will allow the model to be confronted consistently with CMB anisotropies and lensing, weak-lensing measurements, redshift-space distortions, and structure-growth data. The same analysis is required to determine the effective sound speed and to identify possible ghost or gradient instabilities. Big Bang Nucleosynthesis and the thermal history of radiation may provide additional tests of the high-redshift limit. These investigations will determine whether the successful late-time background evolution can be maintained and whether the metric-affine construction can be observationally distinguished from phenomenological interacting-dark-energy and running-vacuum models.

\section{Conclusions}
\label{sect:Conclusion}

We have constructed a cosmological model in which the spin sector of a metric-affine hyperfluid generates an effective interaction between the cosmic fluids and the non-Riemannian geometry. The restricted \(n=1\) realization is unable to accommodate the complete background evolution, whereas the generalized polytropic spin model provides a viable fit to the CC, SNe Ia, and DESI DR2 BAO data considered in this work.  The most general minus branch favors \(n\simeq-1.62\), \(\Omega_{m0}\simeq0.38\), and a phantom-like effective geometric component. Its background fit is competitive with that of \(\Lambda\)CDM, i.e. the AIC and Bayesian evidence \emph{mildly favor} the polytropic model for the combinations containing SNe Ia, while the BIC retains a slight preference for \(\Lambda\)CDM. The present data therefore support the phenomenological viability of the model but do not decisively distinguish between the two cosmological scenarios.

The main result is a background-level \emph{proof of concept} that metric-affine geometry can produce a late-time interacting-fluid cosmology without introducing a direct phenomenological coupling between dark matter and dark energy. Establishing the full viability of this interpretation now requires a perturbative formulation and confrontation with CMB, structure-growth, stability, and early-universe constraints. These extensions will determine whether the geometrical origin of the interaction leads to observable signatures beyond those of existing dynamical-dark-energy and interacting-fluid models.

\acknowledgments
The authors would like to acknowledge the contribution of COST Actions CA21136 ``Addressing observational tensions in cosmology with systematics and fundamental physics (CosmoVerse)''. FA would like to thank Eleonora DiValentino for a number of stimulating conversations. DI's work was supported by the Istituto Nazionale di Fisica Nucleare (INFN), Sezioni di Napoli, {\it Iniziative Specifiche} QGSKY.

\appendix

\bibliographystyle{unsrt}
\bibliography{paperf}
\end{document}